# Optimal Diversity-Multiplexing Tradeoff in Selective-Fading MIMO Channels

*Pedro Coronel and Helmut Bölcskei*

Communication Technology Laboratory

ETH Zurich, 8092 Zurich, Switzerland

E-mail: {pco, boelcskei}@nari.ee.ethz.ch

### Abstract

We establish the optimal diversity-multiplexing (DM) tradeoff of coherent time, frequency, and time-frequency selective-fading multiple-input multiple-output (MIMO) channels and provide a code design criterion for DM tradeoff optimality. Our results are based on the new concept of the "Jensen channel" associated to a given selective-fading MIMO channel. While the original problem seems analytically intractable due to the mutual information between channel input and output being a sum of correlated random variables, the Jensen channel is equivalent to the original channel in the sense of the DM tradeoff and lends itself nicely to analytical treatment. We formulate a systematic procedure for designing DM tradeoff optimal codes for general selective-fading MIMO channels by demonstrating that the design problem can be separated into two simpler and independent problems: the design of an inner code, or precoder, adapted to the channel statistics (i.e., the selectivity characteristics) and an outer code independent of the channel statistics. Our results are supported by appealing geometric intuition, first pointed out for the flat-fading case by Zheng and Tse, *IEEE Trans. Inf. Theory*, 2003.

## I. INTRODUCTION

The diversity-multiplexing (DM) tradeoff framework introduced by Zheng and Tse [1] allows to efficiently characterize the high-SNR rate-reliability tradeoff for communication over multiple-input multiple-output (MIMO) fading channels. The optimal DM tradeoff for flat-fading MIMO channels was characterized in [1]. Sparked by [1] a number of DM tradeoff

Part of this work was performed while the first author was with IBM Research, Zurich Research Laboratory, Switzerland. This work was supported by the STREP project No. IST-026905 MASCOT within the Sixth Framework Programme of the European Commission. This paper was presented in part at the *IEEE Int. Symp. Inf. Theory (ISIT)*, June 2007, Nice, France.

optimal coding/decoding schemes for the flat-fading case were reported during the past few years. In particular, the *non-vanishing* determinant criterion [2], [3] on codeword difference matrices has been shown to constitute a sufficient condition for DM tradeoff optimality [3], [4]; this criterion has led to the construction of DM tradeoff optimal space-time codes based on constellation rotation [3], [5] and cyclic division algebras [4], [6]. Lattice-based space-time codes have been shown to be DM tradeoff optimal in [7]. The DM tradeoff optimality of *approximately universal* space-time codes was established in [8].

*Contributions:* While the results mentioned above focus on frequency-flat block-fading channels, extensions to frequency-selective channels can be found for the single-antenna case in [9], and for the MIMO case in [10]. However, a general characterization of the optimal DM tradeoff in time, frequency, or time-frequency selective-fading MIMO channels, in the following simply referred to as selective-fading MIMO channels, does not seem to be available to date. The present paper resolves this problem for the coherent case[1], provides a code design criterion guaranteeing DM tradeoff optimality, and introduces a systematic procedure for designing DM tradeoff optimal codes. Our results are based on upper and lower bounds on the mutual information of selective-fading MIMO channels; these bounds are shown to exhibit the same DM tradeoff behavior. In particular, we prove that for a given selective-fading MIMO channel the optimal DM tradeoff curve can be obtained by solving the analytically tractable problem of computing the DM tradeoff curve corresponding to its associated "Jensen channel". We demonstrate that the problem of designing DM tradeoff optimal codes can be separated into two simpler and independent problems: the design of an inner code, or precoder, adapted to the channel statistics (i.e., selectivity characteristics) and an outer code independent of the channel statistics. The inner code can be obtained in a systematic fashion as a function of the channel statistics. The design criterion for the outer code is standard with corresponding designs available in the literature.

*Notation:* $M_T$ and $M_R$ denote the number of transmit and receive antennas, respectively. We set $m = \min(M_T, M_R)$ and $M = \max(M_T, M_R)$. For $x \in \mathbb{R}$, we let $[x]^+ = \max(0, x)$. We denote the nonnegative $m$-dimensional orthant by $\mathbb{R}_+^m$. The superscripts $^T$, $^H$, and $^*$ stand for transposition, conjugate transposition, and complex conjugation, respectively. $I_n$ is the $n \times n$ identity matrix, $\mathbb{1}_n$ is the $n \times n$ all ones matrix, $A \otimes B$ and $A \odot B$ denote, respectively, the

---

[1]Throughout the paper, we assume that the receiver has perfect channel state information (CSI) and the transmitter does not have CSI, but is aware of the channel law.

Kronecker and Hadamard products of the matrices $\mathbf{A}$ and $\mathbf{B}$, and $\mathbf{A} \succeq \mathbf{B}$ stands for positive semidefinite ordering. Matrix multiplication has priority over the Kronecker product $\otimes$ and the Hadamard product $\odot$, so that we will write, e.g., $\mathbf{A} \odot \mathbf{BC}$ for $\mathbf{A} \odot (\mathbf{BC})$. $\mathbf{A}^{1/2}$ denotes the (unique) positive semidefinite square root of the positive semidefinite matrix $\mathbf{A}$. For the $n \times m$ matrices $\mathbf{A}_k$ ($k = 1, \ldots, K$), $\mathrm{diag}\{\mathbf{A}_k\}_{k=1}^K$ denotes the $nK \times mK$ block-diagonal matrix with the $k$th diagonal entry given by $\mathbf{A}_k$. If $\mathcal{S}$ is a set, $|\mathcal{S}|$ denotes its cardinality. $\mathbf{A}(\mathcal{S}_1, \mathcal{S}_2)$ stands for the (sub)matrix consisting of the rows of $\mathbf{A}$ indexed by $\mathcal{S}_1$ and the columns of $\mathbf{A}$ indexed by $\mathcal{S}_2$. The columns and rows of the $n \times m$ matrix $\mathbf{A}$ are denoted, respectively, by $\mathbf{a}_k = [\mathbf{A}(1, k) \cdots \mathbf{A}(n, k)]^T$ ($k = 1, \ldots, m$) and $\mathbf{a}_{(p)} = [\mathbf{A}(p, 1) \cdots \mathbf{A}(p, m)]$ ($p = 1, \ldots, n$); $\mathrm{vec}(\mathbf{A}) = [\mathbf{a}_1^T \cdots \mathbf{a}_m^T]^T$. For an $n \times 1$ vector $\mathbf{a} = [a_1 \cdots a_n]^T$, $\mathbf{D}_\mathbf{a} = \mathrm{diag}\{a_m\}_{m=1}^n$, and $\mathbf{a}(m)$ refers to $a_m$. The $n \times n$ FFT matrix $\boldsymbol{\Psi}$ is given by $\boldsymbol{\Psi}(k, l) = \frac{1}{\sqrt{n}} e^{-j\frac{2\pi}{n}(k-1)(l-1)}$ ($k, l = 1, \ldots, n$). The determinant, trace, and rank of $\mathbf{A}$ are denoted as $\det(\mathbf{A})$, $\mathrm{Tr}(\mathbf{A})$, and $\mathrm{rank}(\mathbf{A})$, respectively, and $\|\mathbf{A}\|_\mathrm{F}^2 = \mathrm{Tr}(\mathbf{A}\mathbf{A}^H)$. The nonzero eigenvalues of the $n \times n$ Hermitian matrix $\mathbf{A}$, sorted in ascending order, are designated as $\lambda_k(\mathbf{A})$, $k = 1, \ldots, \mathrm{rank}(\mathbf{A})$. The Kronecker delta function is defined as $\delta_{m,n} = 1$ for $m = n$ and zero otherwise. If $X$ and $Y$ are random variables (RVs), $X \sim Y$ denotes equivalence in distribution, and $\mathbb{E}_X$ is the expectation operator with respect to (w.r.t.) the RV $X$. The random vector $\mathbf{x} \sim \mathcal{CN}(\boldsymbol{\mu}, \mathbf{C})$ is jointly proper Gaussian (JPG) with mean $\boldsymbol{\mu}$ and covariance matrix $\mathbf{C}$. The inner product between two signals $u(t)$ and $v(t)$ is denoted as $\langle u, v \rangle = \int_{-\infty}^{\infty} u(t)v^*(t)dt$. The functions $f(x)$ and $g(x)$ are said to be exponentially equal, denoted by $f(x) \doteq g(x)$, if $\lim_{x \to \infty} \frac{\log f(x)}{\log x} = \lim_{x \to \infty} \frac{\log g(x)}{\log x}$. Exponential inequality, denoted by $\dot{\geq}$ and $\dot{\leq}$, is defined analogously.

## II. Channel and signal model

### A. Channel model

A time-frequency selective single-input single-output (SISO) channel can be modeled as a stochastic linear time-varying (LTV) system [11] with (noise-free) input-output (I/O) relation

$$r(t) = (\mathbb{H}x)(t) = \int_{t'} k_\mathbb{H}(t, t')x(t')dt'$$

where $x(t)$ is the input signal, $r(t)$ is the output signal, and the effect of the channel is described by the linear operator $\mathbb{H}$ with random kernel $k_\mathbb{H}(t, t')$. The time-varying impulse response defined as $h_\mathbb{H}(t, \tau) = k_\mathbb{H}(t, t - \tau)$ yields the equivalent (noise-free) I/O-relation

$$r(t) = \int_{\tau} h_\mathbb{H}(t, \tau)x(t - \tau)d\tau. \tag{1}$$

Two additional system functions that will be important in the ensuing developments are the time-varying transfer function

$$L_{\mathbb{H}}(t, f) = \int_{\tau} h_{\mathbb{H}}(t, \tau) e^{-j2\pi f \tau} d\tau \tag{2}$$

and the spreading function

$$S_{\mathbb{H}}(\tau, \nu) = \int_{t} h_{\mathbb{H}}(t, \tau) e^{-j2\pi \nu t} dt. \tag{3}$$

As an alternative to (1), we may write the I/O-relation in terms of the spreading function as

$$r(t) = \int_{\tau} \int_{\nu} S_{\mathbb{H}}(\tau, \nu) x(t - \tau) e^{j2\pi \nu t} d\tau d\nu. \tag{4}$$

The output signal is thus a weighted superposition of time-frequency shifted replicas of the input signal $x(t)$, where the shifts are parametrized by delay $\tau$ and Doppler shift $\nu$ and $S_{\mathbb{H}}(\tau, \nu)$ corresponds to the weighting function.

*Statistical characterization*: The channel impulse response $h_{\mathbb{H}}(t, \tau)$ is a zero-mean JPG process which is wide-sense stationary in time $t$ and uncorrelated in delay $\tau$, i.e., it satisfies the wide-sense stationary uncorrelated-scattering (WSSUS) assumption [11]

$$\mathbb{E}\{h_{\mathbb{H}}(t, \tau) h_{\mathbb{H}}^{*}(t', \tau')\} = \gamma_{\mathbb{H}}(t - t', \tau) \delta(\tau - \tau').$$

Hence, the time-delay correlation function $\gamma_{\mathbb{H}}(t, \tau)$ fully characterizes the channel statistics. The WSSUS property implies that $L_{\mathbb{H}}(t, f)$ is wide-sense stationary in both $t$ and $f$, and $S_{\mathbb{H}}(\tau, \nu)$ is uncorrelated in delay $\tau$ and Doppler $\nu$:

$$\mathbb{E}\{L_{\mathbb{H}}(t, f) L_{\mathbb{H}}^{*}(t', f')\} = R_{\mathbb{H}}(t - t', f - f')$$

$$\mathbb{E}\{S_{\mathbb{H}}(\tau, \nu) S_{\mathbb{H}}^{*}(\tau', \nu')\} = C_{\mathbb{H}}(\tau, \nu) \delta(\tau - \tau') \delta(\nu - \nu')$$

where the scattering function $C_{\mathbb{H}}(\tau, \nu)$ and the time-frequency correlation function $R_{\mathbb{H}}(t - t', f - f')$ are related through a two-dimensional Fourier transform according to

$$C_{\mathbb{H}}(\tau, \nu) = \int_{t} \int_{f} R_{\mathbb{H}}(t, f) e^{-j2\pi(\nu t - \tau f)} dt \, df. \tag{5}$$

Because $R_{\mathbb{H}}(t, f)$ is stationary in $t$ and $f$, $C_{\mathbb{H}}(\tau, \nu)$ is a real-valued and nonnegative function that can be interpreted as the spectrum of the channel process.

*The underspread assumption and its consequences*: We assume that the channel operator $\mathbb{H}$ is underspread [12] so that the scattering function $C_{\mathbb{H}}(\tau, \nu)$ is compactly supported within the rectangle $[0, \tau_0] \times [0, \nu_0]$, i.e.,

$$C_{\mathbb{H}}(\tau, \nu) = 0 \quad \text{for } (\tau, \nu) \notin [0, \tau_0] \times [0, \nu_0]$$

with the total channel spread $\Delta_{\mathbb{H}} = \tau_0 \nu_0$ satisfying $\Delta_{\mathbb{H}} < 1$. Note that this implies that the spreading function $S_{\mathbb{H}}(\tau, \nu)$ is also supported in this rectangle with probability 1 (w.p.1). The underspread assumption is relevant as most mobile radio channels are (in fact highly) underspread. Moreover, underspread channels have a set of approximate deterministic and structured eigenfunctions which allows to discretize the I/O-relation (4) as described next.

### B. Signaling on approximate eigenfunctions of the channel

We build our developments on the fact that underspread channels are approximately diagonalized by orthogonal Weyl-Heisenberg bases [12] that are obtained by time-frequency shifting a prototype pulse $g(t)$ according to

$$g_{m,k}(t) = g(t - mT)e^{j2\pi kFt}$$

where the grid parameters $T$ and $F$ satisfy $TF \geq 1$ and the basis $\{g_{m,k}(t)\}$ is orthonormal, i.e.,

$$\langle g_{m,k}, g_{n,p} \rangle = \int_t g_{m,k}(t)g_{n,p}^*(t)dt = \delta_{m,n}\delta_{k,p}. \tag{6}$$

Details on the choice of $g(t)$ can be found in [13]. For grid parameters chosen so that $T \leq \frac{1}{\nu_0}$ and $F \leq \frac{1}{\tau_0}$, and hence $TF \leq 1/\Delta_{\mathbb{H}}$, it has been shown in [12], [13] that the impulse response of the underspread fading channel can be well approximated by setting

$$k_{\mathbb{H}}(t, t') = \sum_{m=-\infty}^{\infty} \sum_{k=-\infty}^{\infty} L_{\mathbb{H}}(mT, kF)g_{m,k}(t)g_{m,k}^*(t') \tag{7}$$

where the samples of the time-varying transfer function $L_{\mathbb{H}}(mT, kF)$ are—as a consequence of the assumption on $h_{\mathbb{H}}(t, \tau)$ being a zero-mean JPG process—JPG random variables with zero mean and correlation function

$$\mathbb{E}\{L_{\mathbb{H}}(mT, kF)L_{\mathbb{H}}^*(nT, pF)\} = R_{\mathbb{H}}((m-n)T, (k-p)F). \tag{8}$$

The variance of each channel coefficient $L_{\mathbb{H}}(mT, kF)$ follows from (5) as

$$\sigma_{\mathbb{H}}^2 = \int_\tau \int_\nu C_{\mathbb{H}}(\tau, \nu)d\tau d\nu.$$

*Canonical characterization of signaling schemes*: Based on the developments in the previous paragraph, we construct the transmit signal as a linear combination of the (approximate) eigenfunctions of the channel operator according to

$$x(t) = \sum_{m=-\infty}^{\infty} \sum_{k=0}^{K-1} \tilde{x}_{m,k} g_{m,k}(t) \tag{9}$$

where the $\tilde{x}_{m,k}$ are the information bearing (complex-valued) data symbols. This modulation scheme corresponds to pulse-shaped orthogonal frequency-division multiplexing (OFDM) with symbol duration $T$, tone spacing $F$, and effective signal bandwidth $W = KF$. The receiver computes the inner products $y_{m,k} = \langle y, g_{m,k} \rangle$, where $y(t) = r(t) + z(t)$ and $z(t)$ is additive white Gaussian noise with $\mathbb{E}\{z(t)z^*(t')\} = \delta(t - t')$. Introducing the normalization $x_{m,k} = \frac{1}{\sqrt{\mathsf{SNR}}}\tilde{x}_{m,k}$, with SNR denoting the average signal-to-noise ratio, the overall I/O-relation is given by

$$y_{m,k} = \sqrt{\mathsf{SNR}}\, L_{\mathbb{H}}(mT, kF)x_{m,k} + z_{m,k} \tag{10}$$

where, due to the orthonormality of the basis functions $\{g_{m,k}(t)\}$, the random variables $z_{m,k} = \langle z, g_{m,k} \rangle$ are independent and identically distributed (i.i.d.) across $m$ and $k$, and satisfy $z_{m,k} \sim \mathcal{CN}(0,1)$, for all $m$ and $k$. In essence, this scheme corresponds to transmitting and receiving on the channel's eigenfunctions and, hence, leads to a diagonalization of the channel. For details on the discretization of the I/O-relation (1) described above the interested reader is referred to [13].

## C. Input-output relation with multiple antennas

We assume that communication takes place over $M$ time slots and $K$ frequency slots. For the sake of simplicity of notation, we introduce the bijective mapping $\mathcal{M}$, defined as

$$\begin{aligned} \mathcal{M}: \ \{0, \ldots, M-1\} \times \{0, \ldots, K-1\} &\longrightarrow \{0, \ldots, N-1\} \\ (m,k) &\longmapsto n = mK + k \end{aligned} \tag{11}$$

to index the time-frequency slots $(m,k)$ in (10) according to $n = \mathcal{M}(m,k)$. We extend the I/O-relation (10) to the MIMO case assuming $\mathrm{M_T}$ transmit and $\mathrm{M_R}$ receive antennas, with the scalar subchannels of the $\mathrm{M_R} \times \mathrm{M_T}$ MIMO channel having statistically independent kernels with identical statistics, i.e., with identical scattering functions. Consequently, all subchannels are approximately diagonalized by the same Weyl-Heisenberg basis so that, based on (10) and the mapping in (11), we get

$$\mathbf{y}_n = \sqrt{\frac{\mathsf{SNR}}{\mathrm{M_T}}}\, \mathbf{H}_n \mathbf{x}_n + \mathbf{z}_n, \quad n = 0, \ldots, N-1 \tag{12}$$

where SNR is the average signal-to-noise ratio at each receive antenna, $\mathbf{y}_n$, $\mathbf{x}_n$, and $\mathbf{z}_n$ denote, respectively, the corresponding $\mathrm{M_R} \times 1$ receive signal vector, $\mathrm{M_T} \times 1$ transmit signal vector, and $\mathrm{M_R} \times 1$ JPG noise vector satisfying $\mathbf{z}_n \sim \mathcal{CN}(\mathbf{0}, \mathbf{I}_{\mathrm{M_R}})$, and the channel matrices are given by $\mathbf{H}_n(i,j) = L_{\mathbb{H}}^{(i,j)}(mT, kF)$ $(i = 1, \ldots, \mathrm{M_R},\ j = 1, \ldots, \mathrm{M_T})$, where the superscript $(i,j)$ designates the time-varying transfer function corresponding to the subchannel between transmit antenna $j$ and receive antenna $i$. In the sequel, we shall use $\mathbf{X} = [\mathbf{x}_0 \cdots \mathbf{x}_{N-1}]$ and $\mathbf{Y} = [\mathbf{y}_0 \cdots \mathbf{y}_{N-1}]$ to denote the transmit codeword matrix and the received signal matrix, respectively.

Because the scalar subchannels are assumed to have statistically independent kernels with identical statistics, the channel matrices are spatially uncorrelated and the correlation across slots is given by the time-frequency correlation function in (8). In particular, for any two time-frequency slots $n = \mathcal{M}(m,k)$ and $n' = \mathcal{M}(m',k')$, where $n, n' \in \{0, \ldots, N-1\}$, we have

$$\mathbb{E}\{\mathbf{H}_n(i,j)(\mathbf{H}_{n'}(i,j))^*\} = R_{\mathbb{H}}((m-m')T, (k-k')F) \tag{13}$$

for $i = 1, \ldots, M_R$ and $j = 1, \ldots, M_T$. For later use, we define the corresponding $N \times N$ covariance matrix $\mathbf{R}_{\mathbb{H}}$ as

$$\mathbf{R}_{\mathbb{H}}(n, n') = R_{\mathbb{H}}((m-m')T, (k-k')F) \tag{14}$$

and the stacked channel matrix $\mathbf{H} = [\mathbf{H}_0 \cdots \mathbf{H}_{N-1}]$. Note that with the notation and assumptions in place, we have

$$\mathbb{E}\{\mathrm{vec}(\mathbf{H})(\mathrm{vec}(\mathbf{H}))^H\} = \mathbf{R}_{\mathbb{H}} \otimes \mathbf{I}_{M_T M_R}. \tag{15}$$

The I/O-relation (12) and the channel correlation function (13) are obtained using a signaling scheme that (approximately) diagonalizes the time-frequency selective channel. We stress, however, that (12) is a general I/O-relation that encompasses other widely used models, as for example those in [14], [15, Ch. 3, Sec. 2] used to characterize linear frequency-invariant (LFI) channels and the cyclic signal model resulting from the use of OFDM modulation over linear time-invariant (LTI) channels [16]. The results developed in this paper therefore apply to these models as well provided one takes into account the corresponding structural differences in the covariance matrix (14). We will particularize the main results in this paper to the most important instances of the models used in [14]–[16].

## III. Diversity-multiplexing tradeoff

### A. Preliminaries

When the receiver has perfect CSI, as assumed in this paper, the input distribution that maximizes the mutual information is the Gaussian distribution. Assuming that

$$\mathbb{E}\{\mathrm{vec}(\mathbf{X})(\mathrm{vec}(\mathbf{X}))^H\} = \mathbf{Q}$$

 

where $\mathbf{Q}$ has dimension $N M_\mathrm{T} \times N M_\mathrm{T}$, the maximum mutual information corresponding to the channel in (12) is obtained for $\mathrm{vec}(\mathbf{X}) \sim \mathcal{CN}(\mathbf{0}, \mathbf{Q})$, and is given by

$$I(\mathbf{Y}; \mathbf{X}|\mathbf{D_H}) = \frac{1}{N} \log \det \left( \mathbf{I} + \frac{\mathsf{SNR}}{\mathsf{M_T}} \mathbf{D_H} \mathbf{Q} \, \mathbf{D_H}^H \right) \qquad (16)$$

where $\mathbf{D_H} = \mathrm{diag}\{\mathbf{H}_n\}_{n=0}^{N-1}$. For an average power constraint, specifically $\mathrm{Tr}\,(\mathbf{Q}) \leq N M_\mathrm{T}$, the outage probability at data rate $R$ follows from (16) by optimizing over the input covariance matrix as

$$P_\mathrm{out}(R) = \inf_{\mathbf{Q} \succeq \mathbf{0}, \mathrm{Tr}(\mathbf{Q}) \leq N M_\mathrm{T}} \mathbb{P} \left( \frac{1}{N} \log \det \left( \mathbf{I} + \frac{\mathsf{SNR}}{\mathsf{M_T}} \mathbf{D_H} \mathbf{Q} \, \mathbf{D_H}^H \right) < R \right). \qquad (17)$$

The outage probability is of particular importance for the characterization of the rate-reliability tradeoff because it constitutes a fundamental limit on the error probability. Before proceeding with the analysis of (17), we recall a central concept in the DM tradeoff framework.

A *family of codes* $\mathcal{C}_r$ [1] is a sequence of codebooks $\mathcal{C}_r(\mathsf{SNR})$ parametrized by SNR and with fixed block length. At a given SNR, the corresponding codebook $\mathcal{C}_r(\mathsf{SNR})$ contains $\mathsf{SNR}^{Nr}$ codewords, implying that the data rate $R(\mathsf{SNR})$ scales with SNR according to $R(\mathsf{SNR}) = r \log \mathsf{SNR}$. We say that $\mathcal{C}_r$ operates at multiplexing rate $r \in [0, \mathrm{m}]$. The multiplexing rate $r$ represents the fraction of the ergodic channel capacity that $\mathcal{C}_r$ operates at as SNR increases. The DM tradeoff realized by the family of codes $\mathcal{C}_r$ is characterized by the function

$$d(\mathcal{C}_r) = -\lim_{\mathsf{SNR} \to \infty} \frac{\log P_e(\mathcal{C}_r)}{\log \mathsf{SNR}} \qquad (18)$$

where $P_e(\mathcal{C}_r)$ is the error probability obtained through maximum-likelihood (ML) detection. Moreover, the optimal DM tradeoff curve

$$d^\star(r) = \sup_{\mathcal{C}_r} d(\mathcal{C}_r) \qquad (19)$$

quantifies the maximum achievable diversity gain over all families (w.r.t. SNR) of codes that operate at multiplexing rate $r$.

Following the arguments that lead to [1, Eq. (9)], we shall next show that choosing $\mathbf{Q} = \mathbf{I}$ is DM tradeoff optimal in the selective-fading case as well. More specifically, we demonstrate that $\mathbf{Q} = \mathbf{I}$ solves the optimization problem in (17) in the high-SNR limit. First, we note that an upper bound on $P_\mathrm{out}(R)$ can be obtained by setting $\mathbf{Q} = \mathbf{I}$. On the other hand, because $\mathbf{Q}$ satisfies the power constraint $\mathrm{Tr}\,(\mathbf{Q}) \leq N M_\mathrm{T}$, we necessarily have $\mathbf{Q} \preceq N M_\mathrm{T} \mathbf{I}$. Since $\log \det(\mathbf{A})$ is increasing on the cone of positive definite matrices $\mathbf{A}$ [17, p. 111], replacing $\mathbf{Q}$ by $N M_\mathrm{T} \mathbf{I}$ in

(16) increases the mutual information, and hence yields a lower bound on $P_{\text{out}}(R)$. Combining these arguments, we get

$$\mathbb{P}\left(\frac{1}{N}\sum_{n=0}^{N-1}\log\det\left(\mathbf{I}+\mathsf{SNR}N\mathbf{H}_n\mathbf{H}_n^H\right) < R\right) \leq P_{\text{out}}(R)$$
$$\leq \mathbb{P}\left(\frac{1}{N}\sum_{n=0}^{N-1}\log\det\left(\mathbf{I}+\frac{\mathsf{SNR}}{\mathrm{M_T}}\,\mathbf{H}_n\mathbf{H}_n^H\right) < R\right). \quad (20)$$

Noting that the upper and lower bounds in (20) differ only by a constant factor multiplying the SNR, and using the fact that

$$\lim_{\mathsf{SNR}\to\infty}\frac{\log\mathbb{P}\left(\frac{1}{N}\sum_{n=0}^{N-1}\log\det\left(\mathbf{I}+c\,\mathsf{SNR}\mathbf{H}_n\mathbf{H}_n^H\right) < R\right)}{\log\mathsf{SNR}}$$
$$= \lim_{\mathsf{SNR}\to\infty}\frac{\log\mathbb{P}\left(\frac{1}{N}\sum_{n=0}^{N-1}\log\det\left(\mathbf{I}+c\,\mathsf{SNR}\mathbf{H}_n\mathbf{H}_n^H\right) < R\right)}{\log(c\,\mathsf{SNR})} \quad (21)$$
$$= \lim_{\mathsf{SNR}\to\infty}\frac{\log\mathbb{P}\left(\frac{1}{N}\sum_{n=0}^{N-1}\log\det\left(\mathbf{I}+\mathsf{SNR}\mathbf{H}_n\mathbf{H}_n^H\right) < R\right)}{\log\mathsf{SNR}}$$

for any $c \in \mathbb{R}_+$ independent of SNR, we get

$$P_{\text{out}}(R) \doteq \mathbb{P}(\mathrm{I}(\mathsf{SNR}) < R) \quad (22)$$

where

$$\mathrm{I}(\mathsf{SNR}) = \frac{1}{N}\sum_{n=0}^{N-1}\log\det\left(\mathbf{I}+\frac{\mathsf{SNR}}{\mathrm{M_T}}\,\mathbf{H}_n\mathbf{H}_n^H\right). \quad (23)$$

The outage probability can be characterized in terms of the "singularity levels" of the channel matrices defined as

$$\mu_{n,k} = -\frac{\log\lambda_k(\mathbf{H}_n\mathbf{H}_n^H)}{\log\mathsf{SNR}}, \;\; n = 0,\dots,N-1, \; k = 1,\dots,\mathrm{m}. \quad (24)$$

Rewriting (23) in terms of the singularity levels and letting the data rate scale with SNR as $R(\mathsf{SNR}) = r\log\mathsf{SNR}$, it can be shown by applying [1, Th. 4] that

$$P_{\text{out}}(r\log\mathsf{SNR}) \doteq \mathbb{P}(\mathcal{O}_r) \quad (25)$$

where

$$\mathcal{O}_r = \left\{\boldsymbol{\mu}_n \in \mathbb{R}_+^{\mathrm{m}}, n = 0,\dots,N-1 : \frac{1}{N}\sum_{n=0}^{N-1}\sum_{k=1}^{\mathrm{m}}[1-\mu_{n,k}]^+ < r\right\} \quad (26)$$

with $\boldsymbol{\mu}_n = [\mu_{n,1}\;\cdots\;\mu_{n,\mathrm{m}}]^T$. In the high-SNR limit, the outage probability can be characterized through its SNR exponent given by

$$d_{\mathcal{O}}(r) = -\lim_{\mathsf{SNR}\to\infty}\frac{\log P_{\text{out}}(r\log\mathsf{SNR})}{\log\mathsf{SNR}} = -\lim_{\mathsf{SNR}\to\infty}\frac{\log\mathbb{P}(\mathcal{O}_r)}{\log\mathsf{SNR}} \quad (27)$$

where we used (25). Unlike in the frequency-flat fading case treated in [1], computing $d_{\circ}(r)$ for the selective-fading case seems analytically intractable with the main difficulty stemming from the fact that one has to deal with the sum of correlated (recall that the $\mathbf{H}_n$ are correlated across $n$) terms in (23), for which the joint distribution of the corresponding singularity levels in (24) is in general unknown. It turns out, however, that one can find lower and upper bounds on $\mathrm{I}(\mathsf{SNR})$ in (23) which are exponentially tight in SNR (and, hence, preserve the DM tradeoff behavior) and analytically tractable. The next section formalizes this idea.

Throughout the paper, we shall enforce the peak power constraint

$$\|\mathbf{X}\|_{\mathrm{F}}^2 \le N\mathrm{M_T}, \ \forall \, \mathbf{X} \in \mathcal{C}_r(\mathsf{SNR}). \tag{28}$$

The families of codes $\mathcal{C}_r$ that satisfy the power constraint (28) constitute a subset of the families of codes satisfying the average power constraint induced by $\mathrm{Tr}\,(\mathbf{Q}) \le N M_\mathrm{T}$ and based on which the outage probability in (17) was formulated; it will become manifest, however, that in the high-SNR limit one can find families of codes that satisfy the more restrictive power constraint (28) and still exhibit an error probability that is asymptotically equal to the outage probability. The power constraint (28) implies that the vectorized codeword matrices, i.e., $\mathrm{vec}(\mathbf{X})$, of any (w.r.t. SNR) codebook $\mathcal{C}_r(\mathsf{SNR})$ lie inside a sphere of radius $\sqrt{NM_\mathrm{T}}$ in $\mathbb{C}^{\mathrm{M_T}N}$ centered at the origin. As this sphere radius is constant w.r.t. SNR, its interior becomes increasingly packed with codeword matrices as SNR grows (the codebook size increases according to $|\mathcal{C}_r(\mathsf{SNR})| = \mathsf{SNR}^{Nr}$ to sustain the rate $R(\mathsf{SNR}) = r \log \mathsf{SNR}$). The codeword difference matrices $\mathbf{E} = \mathbf{X} - \mathbf{X}'$, with $\mathbf{X}, \mathbf{X}' \in \mathcal{C}_r(\mathsf{SNR})$, are, therefore, a function of SNR. For the sake of simplicity of notation, we do not make this dependency explicit. In the case $N = 1$ and $\mathrm{M_T} = 1$, for example, an admissible $\mathcal{C}_r$ would be the family of quadrature amplitude modulation (QAM) constellations $\mathcal{A}$ given by

$$\mathcal{A}(\mathsf{SNR}) = \left\{ \sqrt{\frac{2}{\mathsf{SNR}^r}}(a + jb), a, b \in \mathbb{Z} : -\frac{\mathsf{SNR}^{r/2}}{2} \le a, b \le \frac{\mathsf{SNR}^{r/2}}{2} \right\}. \tag{29}$$

Note that $\mathcal{A}(\mathsf{SNR})$ has $|\mathcal{A}(\mathsf{SNR})| = \mathsf{SNR}^r$ constellation points $x$ satisfying the power constraint $x^2 \le 1$. Consequently, the minimum distance in this family of codes scales as[2] $d_{\min}^2 \doteq \mathsf{SNR}^{-r}$, i.e., the area of the unit disk divided by the number of constellation points in $\mathcal{A}(\mathsf{SNR})$.

---

[2] A discussion of the DM tradeoff properties of QAM constellations for the scalar Rayleigh fading channel can be found in [15, Sec. 9.1.2].

*B. Jensen channel and Jensen outage event*

We start by deriving a lower bound on outage probability obtained by upper-bounding the mutual information through Jensen's inequality applied as

$$\mathrm{I}(\mathsf{SNR}) = \frac{1}{N} \sum_{n=0}^{N-1} \log \det \left( \mathbf{I} + \frac{\mathsf{SNR}}{\mathrm{M_T}} \mathbf{H}_n \mathbf{H}_n^H \right) \leq \log \det \left( \mathbf{I} + \frac{\mathsf{SNR}}{\mathrm{M_T} N} \boldsymbol{\mathcal{H}} \boldsymbol{\mathcal{H}}^H \right) \triangleq \mathrm{J}(\mathsf{SNR}) \quad (30)$$

where the "Jensen channel" is an abstract channel characterized by the $\mathrm{m} \times NM$ matrix defined as

$$\boldsymbol{\mathcal{H}} = \begin{cases} [\mathbf{H}_0 \ \cdots \ \mathbf{H}_{N-1}], & \text{if } \mathrm{M_R} \leq \mathrm{M_T}, \\ [\mathbf{H}_0^H \ \cdots \ \mathbf{H}_{N-1}^H], & \text{if } \mathrm{M_R} > \mathrm{M_T}. \end{cases} \quad (31)$$

In the following, we say that a Jensen outage $\mathcal{J}_r$ occurs if the Jensen channel $\boldsymbol{\mathcal{H}}$ is in outage w.r.t. the rate $R = r \log \mathsf{SNR}$, i.e., if $\mathrm{J}(\mathsf{SNR}) < R$. The corresponding outage probability, $P_\mathrm{J}(R) = \mathbb{P}(\mathrm{J}(\mathsf{SNR}) < R)$, clearly satisfies $P_\mathrm{J}(R) \leq P_\mathrm{out}(R)$. The operational significance of the concept of a "Jensen outage" will be established at the end of this section. We shall first focus on characterizing the Jensen outage probability analytically.

Based upon (15), one can show that the Jensen channel can be factored as $\boldsymbol{\mathcal{H}} = \boldsymbol{\mathcal{H}}_w (\mathbf{R}^{T/2} \otimes \mathbf{I_M})$, where $\mathbf{R} = \mathbf{R}_{\mathbb{H}}$, if $\mathrm{M_R} \leq \mathrm{M_T}$, and $\mathbf{R} = \mathbf{R}_{\mathbb{H}}^T$, if $\mathrm{M_R} > \mathrm{M_T}$, and $\boldsymbol{\mathcal{H}}_w$ is the i.i.d. $\mathcal{CN}(0,1)$ matrix with the same dimensions as $\boldsymbol{\mathcal{H}}$ and given by

$$\boldsymbol{\mathcal{H}}_w = \begin{cases} [\mathbf{H}_{w,0} \ \cdots \ \mathbf{H}_{w,N-1}], & \text{if } \mathrm{M_R} \leq \mathrm{M_T}, \\ [\mathbf{H}_{w,0}^H \ \cdots \ \mathbf{H}_{w,N-1}^H], & \text{if } \mathrm{M_R} > \mathrm{M_T}. \end{cases} \quad (32)$$

Here, $\mathbf{H}_{w,n}$ denotes i.i.d. $\mathcal{CN}(0,1)$ matrices of dimension $\mathrm{M_R} \times \mathrm{M_T}$. Using $\boldsymbol{\mathcal{H}}_w \mathbf{U} \sim \boldsymbol{\mathcal{H}}_w$, for any unitary matrix $\mathbf{U}$, and $\lambda_n(\mathbf{R}_{\mathbb{H}}) = \lambda_n(\mathbf{R}_{\mathbb{H}}^T)$ for all $n$, we get $\boldsymbol{\mathcal{H}} \boldsymbol{\mathcal{H}}^H \sim \boldsymbol{\mathcal{H}}_w (\boldsymbol{\Lambda} \otimes \mathbf{I_M}) \boldsymbol{\mathcal{H}}_w^H$, where $\boldsymbol{\Lambda} = \mathrm{diag}\{\lambda_1(\mathbf{R}_{\mathbb{H}}), \ldots, \lambda_\rho(\mathbf{R}_{\mathbb{H}}), 0, \ldots, 0\}$ and we have defined $\rho = \mathrm{rank}(\mathbf{R}_{\mathbb{H}})$. We therefore have

$$\mathrm{J}(\mathsf{SNR}) \sim \log \det \left( \mathbf{I} + \frac{\mathsf{SNR}}{\mathrm{M_T} N} \boldsymbol{\mathcal{H}}_w (\boldsymbol{\Lambda} \otimes \mathbf{I_M}) \boldsymbol{\mathcal{H}}_w^H \right).$$

Next, observe that the following positive semidefinite ordering holds

$$\lambda_1(\mathbf{R}_{\mathbb{H}}) \, \mathrm{diag}\{\mathbf{I}_{\rho M}, \mathbf{0}\} \ \preceq \ \boldsymbol{\Lambda} \otimes \mathbf{I_M} \ \preceq \lambda_\rho(\mathbf{R}_{\mathbb{H}}) \, \mathrm{diag}\{\mathbf{I}_{\rho M}, \mathbf{0}\}. \quad (33)$$

Since, as already noted, $\log \det(\mathbf{A})$ is increasing on the cone of positive definite matrices $\mathbf{A}$ [17, p. 111], we get the following bounds on the Jensen outage probability

$$\mathbb{P}\bigg( \log \det \Big( \mathbf{I} + \lambda_\rho(\mathbf{R}_{\mathbb{H}}) \frac{\mathsf{SNR}}{\mathrm{M_T} N} \overline{\boldsymbol{\mathcal{H}}}_w \overline{\boldsymbol{\mathcal{H}}}_w^H \Big) < R \bigg)$$

$$\leq P_{\mathrm{J}}(R) \tag{34}$$

$$\leq \mathbb{P}\bigg( \log \det \Big( \mathbf{I} + \lambda_1(\mathbf{R}_{\mathbb{H}}) \frac{\mathsf{SNR}}{\mathrm{M_T} N} \overline{\boldsymbol{\mathcal{H}}}_w \overline{\boldsymbol{\mathcal{H}}}_w^H \Big) < R \bigg)$$

where $\overline{\boldsymbol{\mathcal{H}}}_w = \boldsymbol{\mathcal{H}}_w([1\!:\!\mathrm{m}], [1\!:\!\rho\mathrm{M}])$. By the same line of reasoning as in (21), taking the exponential limit (in SNR) in (34) yields

$$P_{\mathrm{J}}(R) \doteq \mathbb{P}\Big( \log \det \Big( \mathbf{I} + \mathsf{SNR}\, \overline{\boldsymbol{\mathcal{H}}}_w \overline{\boldsymbol{\mathcal{H}}}_w^H \Big) < R \Big). \tag{35}$$

The high-SNR asymptotics of $P_{\mathrm{J}}(R)$ can be expressed in terms of the singularity levels of the Jensen channel. Specifically, define $\boldsymbol{\alpha} = [\alpha_1 \; \cdots \; \alpha_{\mathrm{m}}]^T$, where the singularity levels are given by

$$\alpha_k = -\frac{\log \lambda_k(\overline{\boldsymbol{\mathcal{H}}}_w \overline{\boldsymbol{\mathcal{H}}}_w^H)}{\log \mathsf{SNR}}, \quad k = 1, \ldots, \mathrm{m} \tag{36}$$

or, equivalently, $\lambda_k(\overline{\boldsymbol{\mathcal{H}}}_w \overline{\boldsymbol{\mathcal{H}}}_w^H) = \mathsf{SNR}^{-\alpha_k}$. Letting the data rate scale as $R(\mathsf{SNR}) = r \log \mathsf{SNR}$, it can be shown [1, Th. 4] that

$$P_{\mathrm{J}}(r \log \mathsf{SNR}) \doteq \mathbb{P}(\mathcal{J}_r) \tag{37}$$

where

$$\mathcal{J}_r = \left\{ \boldsymbol{\alpha} \in \mathbb{R}_+^{\mathrm{m}} : \alpha_1 \geq \alpha_2 \geq \cdots \geq \alpha_{\mathrm{m}}, \sum_{k=1}^{\mathrm{m}} [1 - \alpha_k]^+ < r \right\}.$$

The corresponding SNR exponent is defined as

$$d_{\mathcal{J}}(r) = -\lim_{\mathsf{SNR} \to \infty} \frac{\log \mathbb{P}(\mathcal{J}_r)}{\log \mathsf{SNR}}.$$

Based on (35), it follows immediately that $d_{\mathcal{J}}(r)$ is nothing but the DM tradeoff curve of an effective MIMO channel with $\rho\mathrm{M}$ transmit and $\mathrm{m}$ receive antennas. We can therefore invoke [1, Th. 2] to infer that the Jensen DM tradeoff curve is the piecewise linear function connecting the points $(r, d_{\mathcal{J}}(r))$ for $r = 0, \ldots, \mathrm{m}$, with

$$d_{\mathcal{J}}(r) = (\rho\mathrm{M} - r)(\mathrm{m} - r). \tag{38}$$

Since, as already noted, $P_{\mathrm{J}}(R) \leq P_{\mathrm{out}}(R)$, it follows that $\mathbb{P}(\mathcal{J}_r) \dot{\leq} \mathbb{P}(\mathcal{O}_r)$. Moreover, by the outage bound [1, Lemma 5], we also get $d^\star(r) \leq d_{\mathcal{O}}(r)$. Hence, in summary, we have

$$d(\mathcal{C}_r) \leq d^\star(r) \leq d_{\mathcal{O}}(r) \leq d_{\mathcal{J}}(r), \quad r \in [0, \mathrm{m}], \tag{39}$$

for any family of codes $\mathcal{C}_r$. The optimal DM tradeoff curve $d^\star(r)$ will be obtained in the next section by deriving a sufficient condition on $\mathcal{C}_r$ to guarantee that $d(\mathcal{C}_r) = d_{\mathcal{J}}(r)$ and hence necessarily $d^\star(r) = d_{\mathcal{J}}(r)$.

## IV. Jensen-optimal code design criterion

The goal of this section is to provide a sufficient condition on a family of codes $\mathcal{C}_r$ to have $d(\mathcal{C}_r) = d_{\mathcal{J}}(r)$. By virtue of (39), this then proves that the optimal DM tradeoff is given by $d_{\mathcal{J}}(r)$ and establishes a design criterion for DM tradeoff optimal codes. Corresponding code constructions are provided in Section V.

### A. Code design criterion

In what follows, for any family of codes $\mathcal{C}_r$, we shall refer to the $N \times N$ matrix $\mathbf{R}_{\mathbb{H}}^T \odot \mathbf{E}^H \mathbf{E}$, where $\mathbf{E} = \mathbf{X} - \mathbf{X}'$ and $\mathbf{X}, \mathbf{X}' \in \mathcal{C}_r(\mathsf{SNR})$, as the *effective codeword difference matrix*. Because the codeword difference matrix $\mathbf{E}$ depends on SNR (see Sec. III-A), so does $\mathbf{R}_{\mathbb{H}}^T \odot \mathbf{E}^H \mathbf{E}$ and any function thereof. In particular, we shall make the SNR-dependency of the eigenvalues of $\mathbf{R}_{\mathbb{H}}^T \odot \mathbf{E}^H \mathbf{E}$ explicit by introducing the notation

$$\lambda_k(\mathsf{SNR}) = \lambda_k(\mathbf{R}_{\mathbb{H}}^T \odot \mathbf{E}^H \mathbf{E}), \ k = 1, \ldots, \rho \mathrm{M_T} \tag{40}$$

where $\lambda_1(\mathsf{SNR}) \leq \lambda_2(\mathsf{SNR}) \leq \cdots \leq \lambda_{\rho \mathrm{M_T}}(\mathsf{SNR})$ for all SNRs.

The following two remarks are in order. First, we note that the remaining $N - \rho \mathrm{M_T}$ eigenvalues of $\mathbf{R}_{\mathbb{H}}^T \odot \mathbf{E}^H \mathbf{E}$ are identically equal to zero for any effective codeword difference matrix arising from $\mathcal{C}_r(\mathsf{SNR})$ and for any SNR. This observation follows from $\mathrm{rank}(\mathbf{A} \odot \mathbf{B}) \leq \mathrm{rank}(\mathbf{A}) \mathrm{rank}(\mathbf{B})$, where $\mathbf{A}$ and $\mathbf{B}$ are positive semidefinite matrices of equal dimensions [18, p. 458]. Since $\mathrm{rank}(\mathbf{R}_{\mathbb{H}}) = \rho$ and $\mathrm{rank}(\mathbf{E}^H \mathbf{E}) \leq \min(\mathrm{M_T}, N) = \mathrm{M_T}$ (recall that $N \geq \rho \mathrm{M_T}$), we have $\mathrm{rank}(\mathbf{R}_{\mathbb{H}}^T \odot \mathbf{E}^H \mathbf{E}) \leq \rho \mathrm{M_T}$, for all $\mathbf{E} = \mathbf{X} - \mathbf{X}'$, $\mathbf{X}, \mathbf{X}' \in \mathcal{C}_r(\mathsf{SNR})$ and all SNRs. In the sequel, we shall refer to the eigenvalues that are not identically equal to zero for all SNR values as *nonzero eigenvalues*.

Second, it is important to note that the eigenvalues $\lambda_k(\mathsf{SNR})$, $k = 1, \ldots, \rho \mathrm{M_T}$, are bounded above by a constant independent of SNR. To see this, note that

$$\lambda_{\rho \mathrm{M_T}}(\mathsf{SNR}) \leq \mathrm{Tr}\left(\mathbf{R}_{\mathbb{H}}^T \odot \mathbf{E}^H \mathbf{E}\right)$$

$$= \sigma_{\mathbb{H}}^2 \mathrm{Tr}\left(\mathbf{E}^H \mathbf{E}\right) \tag{41}$$

$$\leq 4 \sigma_{\mathbb{H}}^2 \mathrm{M_T} N \tag{42}$$

where (41) is a consequence of the fact that the variance of the fading coefficients is $\sigma_{\mathbb{H}}^2$, i.e., the diagonal entries of $\mathbf{R}_{\mathbb{H}}$ are all given by $\sigma_{\mathbb{H}}^2$, and (42) follows from (28) and $\mathbf{E} = \mathbf{X} - \mathbf{X}'$.

Now, (42) is exponentially equal to $\mathsf{SNR}^0 \doteq 1$, which, combined with the ordering imposed on the eigenvalues, shows that

$$\lambda_k(\mathsf{SNR}) \dot{\leq} 1, \ k = 1, \ldots, \rho\mathrm{M_T}. \tag{43}$$

We are now ready to present one of our main results.

*Theorem 1:* Consider a family of codes $\mathcal{C}_r$ with block length $N \geq \rho\mathrm{M_T}$ that operates over the channel (12). For any effective codeword difference matrix, let its eigenvalues be given as in (40), and define

$$\Xi_{\mathrm{m}}^{\rho\mathrm{M_T}}(\mathsf{SNR}) = \min_{\substack{\mathbf{E}=\mathbf{X}-\mathbf{X}', \mathbf{X}\neq\mathbf{X}' \\ \mathbf{X},\mathbf{X}' \in \mathcal{C}_r(\mathsf{SNR})}} \prod_{k=1}^{\mathrm{m}} \lambda_k(\mathsf{SNR}) \tag{44}$$

where the superscript $\rho\mathrm{M_T}$ in $\Xi_{\mathrm{m}}^{\rho\mathrm{M_T}}(\mathsf{SNR})$ emphasizes the fact that there are exactly $\rho\mathrm{M_T}$ nonzero eigenvalues. If $\mathcal{C}_r$ is such that

$$\Xi_{\mathrm{m}}^{\rho\mathrm{M_T}}(\mathsf{SNR}) \dot{\geq} \mathsf{SNR}^{-(r-\epsilon)} \tag{45}$$

for some $\epsilon > 0$ that is constant w.r.t. SNR and $r$, then the corresponding error probability satisfies

$$P_e(\mathcal{C}_r) \doteq \mathsf{SNR}^{-d_{\mathcal{J}}(r)}.$$

*Proof:* Appendix I.

As a direct consequence of Theorem 1, a family of codes $\mathcal{C}_r$ that satisfies (45) realizes a DM tradeoff curve $d(\mathcal{C}_r) = d_{\mathcal{J}}(r)$ and hence, by (39), we obtain

$$d^\star(r) = d_{\mathcal{J}}(r). \tag{46}$$

The optimal DM tradeoff curve for selective-fading MIMO channels is therefore given by the DM tradeoff curve of the associated Jensen channel. Put differently, Theorem 1 shows that, even though $\mathcal{J}_r \subseteq \mathcal{O}_r$ by definition, we still have

$$\mathbb{P}(\mathcal{J}_r) \doteq \mathbb{P}(\mathcal{O}_r)$$

which essentially says that the "original" channel has the same high-SNR outage behavior as its associated Jensen channel. To complete the picture, it remains to show that families of codes satisfying the design criterion (45) indeed exist. This will be done in Section V by providing systematic DM tradeoff optimal code constructions.

## B. Interpretation of the code design criterion

We shall next discuss the relation of the code design criterion (45) to results available in the literature.

*Non-vanishing determinant criterion and approximate universality:* The non-vanishing determinant criterion [2], [3], which is well-known for flat-fading MIMO channels, can be recovered from the code design criterion in Theorem 1 as follows. In the flat-fading case, the channel covariance matrix satisfies $\mathbf{R}_{\mathbb{H}} = \mathbb{1}_N$ with $\rho = 1$, and we hence have $\mathbf{R}_{\mathbb{H}}^T \odot \mathbf{E}^H \mathbf{E} = \mathbf{E}^H \mathbf{E}$ for all possible $\mathbf{E} = \mathbf{X} - \mathbf{X}'$. It follows that the quantity defined in (44) specializes to

$$\Xi_{\mathrm{m}}^{\mathrm{M_T}}(\mathsf{SNR}) = \min_{\substack{\mathbf{E}=\mathbf{X}-\mathbf{X}',\mathbf{X}\neq\mathbf{X}' \\ \mathbf{X},\mathbf{X}' \in \mathcal{C}_r(\mathsf{SNR})}} \prod_{k=1}^{\mathrm{m}} \lambda_k(\mathbf{E}\mathbf{E}^H). \tag{47}$$

For $\mathrm{M_T} \leq \mathrm{M_R}$, we have

$$\Xi_{\mathrm{m}}^{\mathrm{m}}(\mathsf{SNR}) = \min_{\substack{\mathbf{E}=\mathbf{X}-\mathbf{X}',\mathbf{X}\neq\mathbf{X}' \\ \mathbf{X},\mathbf{X}' \in \mathcal{C}_r(\mathsf{SNR})}} \det(\mathbf{E}\mathbf{E}^H)$$

and condition (45) simply requires that $\det(\mathbf{E}\mathbf{E}^H) \dot{\geq} \mathsf{SNR}^{-(r-\epsilon)}$, $\epsilon > 0$, for all codeword difference matrices $\mathbf{E}$. Letting $\tilde{\mathbf{X}} = \sqrt{\mathsf{SNR}^{r/\mathrm{m}}}\, \mathbf{X}$ and $\tilde{\mathbf{E}} = \tilde{\mathbf{X}} - \tilde{\mathbf{X}}'$, it can be readily seen that condition (45) is equivalent to $\det(\tilde{\mathbf{E}}\tilde{\mathbf{E}}^H) \dot{\geq} \mathsf{SNR}^\epsilon$. By taking $\epsilon \to 0$, we get that $\det(\tilde{\mathbf{E}}\tilde{\mathbf{E}}^H)$ must be non-vanishing for increasing SNRs (and hence increasing data rates $R(\mathsf{SNR})$). Examples of code constructions that satisfy the non-vanishing determinant criterion, and which are hence DM tradeoff optimal over i.i.d. Rayleigh flat-fading MIMO channels, can be found in [2]–[6].

The code design criterion of Theorem 1 also encompasses the approximate universality criterion in [8] for flat-fading MIMO channels. This can be seen by specializing (45) to the case $\rho = 1$, i.e.,

$$\Xi_{\mathrm{m}}^{\mathrm{M_T}}(\mathsf{SNR}) \dot{\geq} \mathsf{SNR}^{-(r-\epsilon)}, \ \epsilon > 0 \tag{48}$$

and comparing (48) to the criterion given in [8, Theorem 3.1]. The coincidence of the approximate universality criterion and (45) (in flat fading) is noteworthy as the criteria are arrived at using completely different assumptions and different corresponding proof techniques: While our result is based on explicit assumptions on the channel fading statistics, the approximate universality condition guarantees DM tradeoff optimal performance for every fading distribution, over any channel that is not in outage.

*Relation to classical space-time code design criteria:* Next, we specialize our code design criterion to multiplexing rate $r = 0$, i.e., the data rate is fixed and does not increase with SNR, in which case the same codebook can be used for all SNR values. Note that this implies that the eigenvalues in (40) are no longer functions of SNR. From Theorem 1, it follows that the codebook is DM tradeoff optimal if it satisfies $\Xi_m^{\rho M_T}(\mathsf{SNR}) \dot{\geq} \mathsf{SNR}^\epsilon$, $\epsilon > 0$, or, equivalently, if every effective codeword difference matrix $\mathbf{R}_{\boxplus}^T \odot \mathbf{E}^H \mathbf{E}$ has $\rho M_T$ nonzero eigenvalues. This is to say that the sufficient condition for DM tradeoff optimality at $r = 0$ can be stated as

$$\text{rank}\big(\mathbf{R}_{\boxplus}^T \odot \mathbf{E}^H \mathbf{E}\big) = \rho M_T, \ \forall \mathbf{E} = \mathbf{X} - \mathbf{X}', \ \mathbf{X} \neq \mathbf{X}', \ \mathbf{X}, \mathbf{X}' \in \mathcal{C}_r. \tag{49}$$

This is precisely the code design criterion found in the SISO case in [19] using the same channel model as here and in [20] in the context of MIMO-OFDM modulation.

### C. Geometric interpretation of the optimal DM tradeoff

In the following, we provide a geometric interpretation of the optimal DM tradeoff. The discussion follows closely the corresponding analysis for the flat-fading case reported in [1]. To simplify the exposition, we consider the case of OFDM modulation over ISI channels and start by noting that in an OFDM system with $N$ tones the I/O-relation (after discarding the cyclic prefix at the receiver) is given by (12) with

$$\mathbf{H}_n = \sum_{l=0}^{L-1} \mathbf{H}(l) e^{-j\frac{2\pi}{N}ln} \tag{50}$$

where $\mathbf{H}(l), l = 0, \ldots, L-1$, denotes the i.i.d. matrix-valued channel taps with $\mathcal{CN}(0, 1)$ entries. The corresponding mutual information (23) can thus be written as

$$\mathrm{I}(\mathsf{SNR}) = \frac{1}{N} \log \det \left( \mathbf{I} + \frac{\mathsf{SNR}}{M_T} \mathbf{D_H} \mathbf{D_H}^H \right)$$

where we recall that $\mathbf{D_H} = \text{diag}\{\mathbf{H}_n\}_{n=0}^{N-1}$. Following the geometric argument in the flat-fading case [1], we wish to relate the outage probability at multiplexing rate $r$ to the rank of the matrix $\mathbf{D_H}$. Unfortunately, $\text{rank}(\mathbf{D_H})$ is difficult to characterize, in general, because the corresponding diagonal blocks are correlated due to (50). In an OFDM system, the matrix $\mathbf{D_H} \mathbf{D_H}^H$ can, however, readily be shown to be unitarily equivalent to $\mathbf{C_H} \mathbf{C_H}^H$, where $\mathbf{C_H}$ is the following $N M_R \times N M_T$

block-circulant matrix

$$\mathbf{C_H} = \begin{bmatrix} \mathbf{H}(0) & \mathbf{0} & \cdots & \cdots & \mathbf{0} & \mathbf{H}(L-1) & \cdots & \mathbf{H}(1) \\ \mathbf{H}(1) & \mathbf{H}(0) & \ddots & & \vdots & \mathbf{0} & \ddots & \vdots \\ \vdots & \mathbf{H}(1) & \ddots & \ddots & \vdots & \vdots & \ddots & \mathbf{H}(L-1) \\ \mathbf{H}(L-1) & \vdots & \ddots & \ddots & \mathbf{0} & \vdots & & \mathbf{0} \\ \mathbf{0} & \mathbf{H}(L-1) & & \ddots & \mathbf{H}(0) & \mathbf{0} & & \vdots \\ \vdots & \mathbf{0} & \ddots & & \mathbf{H}(1) & \ddots & \ddots & \vdots \\ \vdots & & \ddots & \ddots & \vdots & \ddots & \ddots & \mathbf{0} \\ \mathbf{0} & \cdots & \cdots & \mathbf{0} & \mathbf{H}(L-1) & \cdots & \mathbf{H}(1) & \mathbf{H}(0) \end{bmatrix}.$$

For $N > L$ (which is satisfied in any OFDM system), the structure of $\mathbf{C_H}$ implies that its rank is completely determined by the rank of its first $M_T$ columns in the case $M_T \le M_R$ and by the rank of its last $M_R$ rows in the case $M_R < M_T$. More specifically, $\mathrm{rank}(\mathbf{C_H})$ satisfies (for every channel realization)

$$\mathrm{rank}(\mathbf{C_H}) = N\,\mathrm{rank}(\mathbf{C}_w) \tag{51}$$

where

$$\mathbf{C}_w = \begin{cases} \mathbf{C_H}([1:LM_R],[1:M_T])^T, & \text{if } M_T \le M_R \\ \mathbf{C_H}([(N-1)M_R+1:NM_R],[(N-L)M_T+1:NM_T]), & \text{if } M_T > M_R. \end{cases}$$

Note that $\mathbf{C}_w$ is an $m \times LM$ matrix with i.i.d. $\mathcal{CN}(0,1)$ entries and that it is equal in distribution to $\overline{\mathcal{H}}_w$ (cf. (105) and (108)) obtained from the Jensen channel. In order to characterize $\mathrm{rank}(\mathbf{C_H})$, it follows from (51) that it suffices to characterize $\mathrm{rank}(\mathbf{C}_w)$. In particular, following [1], we shall be interested in determining the number of parameters required to specify a matrix $\mathbf{C_H}$ of rank $Nr$, or, equivalently, a matrix $\mathbf{C}_w$ of rank $r$. This number is obtained as follows: $LMr$ parameters are required to specify $r$ linearly independent rows in $\mathbf{C}_w$. The remaining $m - r$ rows are then given by linear combinations of these $r$ linearly independent rows. Specifying these linearly dependent rows requires $r$ parameters per row (i.e., the coefficients in the linear combinations of the $r$ linearly independent rows) and hence $(m - r)r$ parameters overall. The total number of parameters specifying a matrix $\mathbf{C_H}$ of rank $Nr$ is therefore obtained as

$$LMr + (m-r)r = LMm - (LM-r)(m-r). \tag{52}$$

Now, following the reasoning in [1, Sec. 3.2], we can conclude that an outage at multiplexing rate $r$ occurs when $\mathbf{C}_w$ is close to the manifold of all rank-$r$ matrices. This requires a collapse in the components of $\mathbf{C}_w$ in all the dimensions[3] orthogonal to that subspace; the number of such dimensions is given by $(LM - r)(\mathrm{m} - r)$, which is precisely the SNR exponent given in (38) and hence concludes the argument.

### D. Particularizing the design criterion (49) to ISI channels

Condition (49) can be stated in a form that yields geometric insight into the code design problem and nicely reveals the code design criterion reported in [20] for frequency-selective MIMO channels as a special case. We start by stating the following result in full generality and will then specialize it to the case of ISI channels.

*Proposition 1:* Let $\mathbf{R}_{\mathbb{H}} = \sum_{n=0}^{\rho-1} \lambda_n \mathbf{u}_n \mathbf{u}_n^H$ be the eigenvalue decomposition of the channel covariance matrix. Then, (49) holds if and only if

$$\boldsymbol{\Delta} = \left[ \sqrt{\lambda_0} \, \mathbf{D}_{\mathbf{u}_0^*} \mathbf{E}^H \ \cdots \ \sqrt{\lambda_{\rho-1}} \, \mathbf{D}_{\mathbf{u}_{\rho-1}^*} \mathbf{E}^H \right]^H \tag{53}$$

has full rank.

*Proof:* Based on the eigenvalue decomposition of $\mathbf{R}_{\mathbb{H}}$, we get

$$\mathbf{R}_{\mathbb{H}}^T \odot \mathbf{E}^H \mathbf{E} = \left( \sum_{n=0}^{\rho-1} \lambda_n \mathbf{u}_n^* \mathbf{u}_n^T \right) \odot \mathbf{E}^H \mathbf{E}$$

$$= \sum_{n=0}^{\rho-1} \lambda_n \mathbf{D}_{\mathbf{u}_n^*} \mathbf{E}^H \mathbf{E} \, \mathbf{D}_{\mathbf{u}_n} \tag{54}$$

$$= \boldsymbol{\Delta}^H \boldsymbol{\Delta} \tag{55}$$

where (54) follows from the fact that $\mathbf{a}\mathbf{b}^T \odot \mathbf{C} = \mathbf{D}_{\mathbf{a}} \mathbf{C} \mathbf{D}_{\mathbf{b}}$ for any $n \times 1$ vectors $\mathbf{a}, \mathbf{b}$ and any $n \times n$ matrix $\mathbf{C}$. The proof is concluded upon noting that $\mathrm{rank}(\boldsymbol{\Delta}) = \mathrm{rank}(\boldsymbol{\Delta}^H \boldsymbol{\Delta})$. ∎

We note that a decomposition of the effective codeword difference matrix similar to that in (53) has also been reported for the SISO case in [19].

*Specialization to the ISI channel case:* We shall next specialize Proposition 1 to the ISI channel case, and recover the code design criterion reported in [20] for MIMO ISI channels. In an OFDM system, as considered in [20], the channel's covariance matrix is given by

$$\mathbf{R}_{\mathbb{H}} = \boldsymbol{\Psi} \, \mathrm{diag}\left\{ \sigma_0^2, \ldots, \sigma_{L-1}^2, 0, \ldots, 0 \right\} \boldsymbol{\Psi}^H \tag{56}$$

where the $\{\sigma_l^2\}$ correspond to the power-delay profile that, for the sake of simplicity of exposition, we assume to be given by $\sigma_l^2 = 1$, for all $l$, throughout this section. Since $\mathbf{R}_{\mathbb{H}}$ is diagonalized by the FFT matrix $\mathbf{\Psi}$, we have $\mathbf{D}_{\mathbf{u}_n^*} = \mathbf{D}^n$, where $\mathbf{D} = \frac{1}{\sqrt{N}}\mathrm{diag}\{e^{j\frac{2\pi}{N}k}\}_{k=0}^{N-1}$. Hence, based on (56), (53) specializes to [20]

$$\mathbf{\Delta} = \left[\mathbf{D}^0\mathbf{E}^H \ \cdots \ \mathbf{D}^{L-1}\mathbf{E}^H\right]^H.$$

---

[3]We refer to [15, note on p. 397] for an argument on why it is meaningful to talk about orthogonal dimensions even though the manifold of all rank-$r$ matrices is not a linear subspace.

Since the rank of a matrix is unaltered by left multiplication by a full-rank matrix, we can equivalently consider the matrix $\mathbf{\Psi}^T\mathbf{\Delta}$. In particular, we note that $\mathbf{\Psi}^T\mathbf{D}^n\mathbf{E}^H = \mathbf{\Pi}^n\mathbf{E}_t^H$, where $\mathbf{\Pi} = [\boldsymbol{\pi}_1 \ \cdots \ \boldsymbol{\pi}_{N-1} \ \boldsymbol{\pi}_0]$, with $\boldsymbol{\pi}_k(n) = 1$ for $k = n$ and $\boldsymbol{\pi}_k(n) = 0$ otherwise, is the basic circulant permutation matrix and $\mathbf{E}_t = \mathbf{E}\mathbf{\Psi}^*$ is a time-domain representation of the codeword difference matrix. The code design criterion for $r = 0$ in the ISI case therefore amounts to ensuring that the matrix

$$\left[\mathbf{\Pi}^0\mathbf{E}_t^H \ \cdots \ \mathbf{\Pi}^{L-1}\mathbf{E}_t^H\right]^H \tag{57}$$

has full rank for all codeword difference matrices, which is precisely the code design criterion reported in [20], [21]. Requiring the matrix in (57) to have full rank for all $\mathbf{E}_t$ essentially amounts to saying that the code should be designed such that the receiver can separate the shifted versions of the transmit signal.

*Prior results on the DM tradeoff for ISI channels:* We shall next specialize our results to frequency-selective fading MIMO channels, recovering the results reported previously in [9], [10]. Assuming a frequency-selective fading channel with $L$ taps that are i.i.d. $\mathcal{CN}(0, 1)$ and a cyclic I/O-relation (as in an OFDM system), the covariance matrix is again given by (56) with $\rho = \mathrm{rank}(\mathbf{R}_{\mathbb{H}}) = L$. Inserting $\rho = L$ into (38) and using (46) yields the optimal DM tradeoff curve as the piecewise linear function connecting the points $(r, d^\star(r))$ for $r = 0, \ldots, \mathrm{m}$, with

$$d^\star(r) = (L\mathrm{M} - r)(\mathrm{m} - r). \tag{58}$$

This is the optimal DM tradeoff curve for frequency-selective fading MIMO channels reported previously in [10]. Specializing (58) to the single-antenna case $\mathrm{M_T} = \mathrm{M_R} = 1$ and noting that $d^\star(r) = (L - r)(1 - r) = L(1 - r)$ for $r = 0, 1$, recovers the result reported in [9]. We note that the proof techniques employed in [9], [10] are different from the approach taken in this paper and seem to be tailored to the frequency-selective case. In addition, since Theorem 1 only requires $N \geq L\mathrm{M_T}$, our result is not limited to large block lengths as required in [9], [10].

Finally, we note that the achievable DM tradeoff curve reported in [1] for the case where coding is performed across $L$ independent MIMO channels is given by

$$d_{\mathrm{I}}(r) = L(\mathrm{M} - r)(\mathrm{m} - r).$$

We clearly have $d_1(r) \leq d^\star(r)$ for all multiplexing rates and all possible values of $\mathrm{m}$ and $\mathrm{M}$.

*The case of linear convolution:* For linear convolution, as encountered in single-carrier modulation, the code design criterion for $r = 0$ is obtained by replacing $\mathbf{\Pi}$ in (57) by the forward shift matrix [18] and ensuring that the resulting matrix has full rank for all codeword difference matrices. To see this, consider the following I/O-relation

$$\mathbf{y}[n] = \sqrt{\frac{\mathsf{SNR}}{\mathrm{M_T}}} \sum_{l=0}^{L-1} \mathbf{H}(l)\, \mathbf{x}[n-l] + \mathbf{z}[n] \tag{59}$$

where $\mathbf{y}[n]$, $\mathbf{x}[n]$, and $\mathbf{z}[n]$ denote the received, transmitted, and noise vector sequences, respectively. We assume that $\mathbf{x}[n] = \mathbf{0}$ for $n < 0$ and $n > N - L$, and consider the time interval $n = 0, \ldots, N-1$. Stacking the received signal vectors according to $\mathbf{Y} = [\mathbf{y}[0] \; \cdots \; \mathbf{y}[N-1]]$ and the channel taps as $\mathbf{H} = [\mathbf{H}(0) \; \cdots \; \mathbf{H}(L-1)]$, the resulting I/O-relation can be written as

$$\mathbf{Y} = \sqrt{\frac{\mathsf{SNR}}{\mathrm{M_T}}}\, \mathbf{H} \boldsymbol{\mathcal{X}} + \mathbf{Z} \tag{60}$$

where $\mathbf{Z} = [\mathbf{z}[0] \; \cdots \; \mathbf{z}[N-1]]$ and the $L\mathrm{M_T} \times N$ transmit signal matrix is given by

$$\boldsymbol{\mathcal{X}} = \begin{bmatrix} \mathbf{x}[0] & \mathbf{x}[1] & \cdots & \mathbf{x}[N-L] & 0 & \cdots & 0 \\ 0 & \mathbf{x}[0] & \mathbf{x}[1] & \cdots & \mathbf{x}[N-L] & \ddots & \vdots \\ \vdots & \ddots & \ddots & \ddots & & \ddots & 0 \\ 0 & \cdots & 0 & \mathbf{x}[0] & \mathbf{x}[1] & \cdots & \mathbf{x}[N-L] \end{bmatrix}.$$

Consequently, any codeword difference matrix $\boldsymbol{\mathcal{E}} = \boldsymbol{\mathcal{X}} - \boldsymbol{\mathcal{X}}'$ has the structure

$$\boldsymbol{\mathcal{E}} = \begin{bmatrix} \mathbf{S}^0 \mathbf{E}^H & \cdots & \mathbf{S}^{L-1} \mathbf{E}^H \end{bmatrix}^H \tag{61}$$

where $\mathbf{S}$ denotes the forward shift matrix and, here, $\mathbf{E} = [\mathbf{e}[0] \; \cdots \; \mathbf{e}[N-L+1] \; \mathbf{0} \cdots \mathbf{0}]$ with $\mathbf{e}[n] = \mathbf{x}[n] - \mathbf{x}'[n]$. Comparing (61) with (57) shows that the code design criterion follows from (57) by replacing the cyclic shifts by linear shifts, and ensuring full-rank of the resulting codeword difference matrices [20].

## E. Block-fading channels

In the block-fading channel model, the channel remains unchanged during a block of say $L$ time slots and changes in a statistically independent fashion across blocks. We consider $B$ such independent blocks for which the I/O-relation (12) holds with $N = BL$ and

$$\mathbf{H}_n = \mathbf{H}\left( \left\lfloor \frac{n}{L} \right\rfloor + 1 \right), \; n = 0, \ldots, N-1$$

where $\mathbf{H}(b), b = 1, \ldots, B$, denotes the channel matrix with i.i.d. $\mathcal{CN}(0,1)$ entries corresponding to the $b$th block. The $BL \times BL$ channel covariance matrix $\mathbf{R}_{BF}$ is therefore given by

$$\mathbf{R}_{BF} = \mathbf{I}_B \otimes \mathbb{1}_L$$

with $\mathrm{rank}(\mathbf{R}_{BF}) = B$. The corresponding Jensen DM tradeoff curve is the piecewise linear function connecting the points $(r, d_{\mathcal{J}}(r))$ for $r = 0, \ldots, \mathrm{m}$, where $d_{\mathcal{J}}(r) = (BM - r)(\mathrm{m} - r)$.

Theorem 1 provides a sufficient condition for a family of codes $\mathcal{C}_r$ with block length $N \geq BM_T$ to achieve the optimal DM tradeoff curve. In the block-fading case, every codeword $\mathbf{X} \in \mathcal{C}_r(\mathsf{SNR})$ can be partitioned into $B$ blocks of size $M_R \times L$ according to $\mathbf{X} = [\mathbf{X}_1 \cdots \mathbf{X}_B]$ and, similarly, any codeword difference matrix $\mathbf{E} = \mathbf{X} - \mathbf{X}'$ can be represented as $\mathbf{E} = [\mathbf{E}_1 \cdots \mathbf{E}_B]$, where $\mathbf{E}_b = \mathbf{X}_b - \mathbf{X}'_b$, for $b = 1, \ldots, B$, has dimension $M_R \times L$. Consequently, the effective codeword difference matrices have the following structure:

$$\mathbf{R}_{BF}^T \odot \mathbf{E}^H \mathbf{E} = \mathrm{diag}\big\{\mathbf{E}_b^H \mathbf{E}_b\big\}_{b=1}^B$$

and the corresponding code design criterion follows from (45) as

$$\prod_{k=1}^{\mathrm{m}} \lambda_k(\mathbf{R}_{BF}^T \odot \mathbf{E}^H \mathbf{E}) \,\dot{\geq}\, \mathsf{SNR}^{-(r-\epsilon)} \tag{62}$$

for all possible codeword difference matrices $\mathbf{E}$ arising from $\mathcal{C}_r(\mathsf{SNR})$, and some $\epsilon > 0$ constant w.r.t. $\mathsf{SNR}$ and $r$. We note that the block diagonal structure of the effective codeword difference matrices implies that

$$\Big\{\lambda_1(\mathbf{R}_{BF}^T \odot \mathbf{E}^H \mathbf{E}), \ldots, \lambda_{BM_T}(\mathbf{R}_{BF}^T \odot \mathbf{E}^H \mathbf{E}), \underbrace{0, \ldots, 0}_{N-BM_T}\Big\}$$

$$= \bigcup_{b=1}^B \Big\{\lambda_1(\mathbf{E}_b^H \mathbf{E}_b), \ldots, \lambda_{M_T}(\mathbf{E}_b^H \mathbf{E}_b), \underbrace{0, \ldots, 0}_{L-M_T}\Big\}. \tag{63}$$

In the absence of coding across individual blocks, that is, if the codewords are designed so that they satisfy the following per-block criteria obtained from (45)

$$\prod_{l=1}^{\mathrm{m}} \lambda_l(\mathbf{E}_b^H \mathbf{E}_b) \,\dot{\geq}\, \mathsf{SNR}^{-(r-\epsilon)}, \; \epsilon > 0, \; \text{for } b = 1, \ldots, B, \tag{64}$$

the design criterion (62) is not guaranteed to be satisfied because the $\mathrm{m}$ smallest nonzero[4] eigenvalues of $\mathbf{R}_{BF}^T \odot \mathbf{E}^H \mathbf{E}$ are, in general, not equal to the $\mathrm{m}$ smallest nonzero eigenvalues of $\mathbf{E}_{b'}^H \mathbf{E}_{b'}$ for some $b' \in \{1, \ldots, B\}$. We can therefore conclude that having the individual blocks

---

[4]Recall that "nonzero eigenvalue" refers to an eigenvalue that is not identically equal to zero for all SNR values.

 

$\mathbf{E}_b$ satisfy (64) is, in general, not sufficient to ensure DM tradeoff optimality and coding across blocks is required.

Interestingly, the situation is different for $\mathrm{M_T} = 1$. In this case, we have $\mathrm{m} = 1$ so that (62) is given by

$$\lambda_1(\mathbf{R}_{BF}^T \odot \mathbf{E}^H \mathbf{E}) \,\dot{\geq}\, \mathsf{SNR}^{-(r-\epsilon)}, \ \epsilon > 0. \tag{65}$$

We also note that there is only one nonzero eigenvalue per block, and the per-block design criterion in (64) now reads

$$\lambda_1(\mathbf{E}_b^H \mathbf{E}_b) \,\dot{\geq}\, \mathsf{SNR}^{-(r-\epsilon)}, \ \epsilon > 0, \ \text{for } b = 1, \dots, B. \tag{66}$$

Since $\lambda_1(\mathbf{R}_{BF}^T \odot \mathbf{E}^H \mathbf{E}) = \lambda_1(\mathbf{E}_{b'}^H \mathbf{E}_{b'})$ for some $b' \in \{1, \dots, B\}$, we can conclude that satisfying (66) for all blocks guarantees that (65) is also satisfied.

## V. Code design for optimal DM tradeoff

We established the optimal DM tradeoff for the general class of selective-fading channels and provided a code design criterion for achieving DM tradeoff optimality. The goal of this section is to demonstrate the existence of codes satisfying this design criterion and to provide corresponding systematic design procedures. In addition, we want to ensure that the proposed DM tradeoff optimal code designs are practicable in the sense of being independent of the channel covariance matrix (i.e., of the selectivity characteristics). We shall see that in the single transmit antenna case this is rather straightforward to accomplish. In the case of multiple transmit antennas, we propose a procedure that decouples the problem into the design of a precoder (which can be obtained systematically for a given $\mathbf{R}_{\mathbb{H}}$) and an outer code which has to satisfy a design criterion that is independent of $\mathbf{R}_{\mathbb{H}}$.

### A. The single transmit antenna case

Consider the case $\mathrm{M_T} = 1$ and $\mathrm{M_R}$ general with a corresponding family of codes $\mathcal{C}_r$ of block length $N$. The codewords in $\mathcal{C}_r$ are $1 \times N$ vectors of the form $\mathbf{x} = [x_0 \ \cdots \ x_{N-1}]$ with the corresponding effective codeword difference matrices given by

$$\mathbf{R}_{\mathbb{H}}^T \odot \mathbf{e}^H \mathbf{e} = \mathbf{D}_{\mathbf{e}}^H \mathbf{R}_{\mathbb{H}}^T \mathbf{D}_{\mathbf{e}} \tag{67}$$

so that $\Xi_1^\rho(\mathsf{SNR})$ defined in (44) specializes to

$$\Xi_1^\rho(\mathsf{SNR}) = \min_{\substack{\mathbf{e}=\mathbf{x}-\mathbf{x}', \mathbf{x} \neq \mathbf{x}' \\ \mathbf{x},\mathbf{x}' \in \mathcal{C}_r(\mathsf{SNR})}} \lambda_1(\mathbf{D}_{\mathbf{e}}^H \mathbf{R}_{\mathbb{H}}^T \mathbf{D}_{\mathbf{e}}). \tag{68}$$

 

The dependency of (68) on $\mathbf{R}_{\mathbb{H}}$ leads to different code design criteria depending on the channel selectivity characteristics. For example, in a flat-fading channel, where $\mathbf{R}_{\mathbb{H}} = \mathbb{1}_N$, $\rho = 1$, and $\mathbf{R}_{\mathbb{H}}^T \odot \mathbf{e}^H \mathbf{e} = \mathbf{e}^H \mathbf{e}$, we have $\Xi_1^1(\mathsf{SNR}) = \min_{\mathbf{e} \neq \mathbf{0}} \|\mathbf{e}\|^2$. On the other hand, in the fast-fading case where $\mathbf{R}_{\mathbb{H}} = \mathbf{I}$ and hence $\rho = N$, it follows from (68) that

$$\Xi_1^N(\mathsf{SNR}) = \min_{\substack{\mathbf{e} \neq \mathbf{0} \\ n = 0, \ldots, N-1}} |e_n|^2.$$

We shall next provide a code design criterion which guarantees DM tradeoff optimality irrespectively of $\mathbf{R}_{\mathbb{H}}$.

*Proposition 2:* The family of codes $\mathcal{C}_r$ is DM tradeoff optimal for $\mathrm{M_T} = 1$ if it satisfies

$$\min_{\substack{\mathbf{e} = \mathbf{x} - \mathbf{x}', \mathbf{x} \neq \mathbf{x}' \\ \mathbf{x}, \mathbf{x}' \in \mathcal{C}_r(\mathsf{SNR})}} \min_n |e_n|^2 \gtrsim \mathsf{SNR}^{-(r-\epsilon)} \tag{69}$$

for some $\epsilon > 0$ constant w.r.t. $\mathsf{SNR}$ and $r$.

*Proof:* Applying Ostrowski's Theorem [18, Theorem 4.5.9] to the effective codeword difference matrix (67) and using $\lambda_k(\mathbf{R}_{\mathbb{H}}^T) = \lambda_k(\mathbf{R}_{\mathbb{H}})$ yields $\lambda_n(\mathbf{D}_{\mathbf{e}}^H \mathbf{R}_{\mathbb{H}}^T \mathbf{D}_{\mathbf{e}}) = \theta_{\mathbf{e}} \lambda_n(\mathbf{R}_{\mathbb{H}})$, $n = 0, \ldots, N-1$, where $\theta_{\mathbf{e}} \in [\min_n |e_n|^2, \max_n |e_n|^2]$. Hence, by (69), we have

$$\lambda_k(\mathbf{D}_{\mathbf{e}}^H \mathbf{R}_{\mathbb{H}}^T \mathbf{D}_{\mathbf{e}}) \gtrsim \mathsf{SNR}^{-(r-\epsilon)} \lambda_k(\mathbf{R}_{\mathbb{H}}), \quad k = 0, \ldots, \rho-1, \tag{70}$$

for all $\mathbf{e} \neq \mathbf{0}$. Since the eigenvalues of $\mathbf{R}_{\mathbb{H}}$ are constant w.r.t. $\mathsf{SNR}$, we conclude from (70) that $\Xi_1^\rho(\mathsf{SNR}) \gtrsim \mathsf{SNR}^{-(r-\epsilon)}$, implying by (45) that $\mathcal{C}_r$ is DM tradeoff optimal. ∎

Since the minimum distance in a QAM constellation scales as $d_{\min}^2 \doteq \mathsf{SNR}^{-r}$ [15, Sec. 9.1.2], using uncoded QAM constellations with $\mathsf{SNR}^r$ points in each slot $n = 0, \ldots, N-1$ satisfies (69) for $\epsilon \to 0$. We can therefore conclude from Proposition 2 that in the single transmit antenna case uncoded QAM is DM tradeoff optimal irrespectively of $\mathbf{R}_{\mathbb{H}}$.

## B. Multiple transmit antennas

For multiple transmit antennas, the situation is more complicated. We next describe a procedure that decouples the problem of designing DM tradeoff optimal codes for multiple transmit antennas into the design of a precoder depending on $\mathbf{R}_{\mathbb{H}}$ and an outer code which has to satisfy a design criterion that is independent of $\mathbf{R}_{\mathbb{H}}$. Specifically, we shall see that the precoder can be chosen such that the criterion to be satisfied by the outer code boils down to a criterion well-known in the literature with corresponding optimal code designs available.

We consider families (w.r.t. SNR) of codes of block length $N$ for which the $\mathrm{M_T} \times N$ codeword matrices are given by

$$\tilde{\mathbf{X}} = \mathbf{P} \odot \mathbf{X}. \tag{71}$$

The matrix $\mathbf{P}$ can be thought of as an inner code, or precoder, and $\mathbf{X}$ can be interpreted as a codeword matrix belonging to an outer family of codes $\mathcal{C}_r$. In what follows, we shall refer to $\mathcal{C}_r$ simply as a family of codes.

If $\mathbf{X}, \mathbf{X}' \in \mathcal{C}_r(\mathsf{SNR})$, the corresponding precoded codeword difference matrix is given by $\tilde{\mathbf{E}} = \mathbf{P} \odot \mathbf{E}$, where $\mathbf{E} = \mathbf{X} - \mathbf{X}'$. With the rows of $\mathbf{E}$ and $\mathbf{P}$ denoted as $\mathbf{e}_{(l)}$ and $\mathbf{p}_{(l)}$, respectively, we have

$$\tilde{\mathbf{E}}^H \tilde{\mathbf{E}} = \sum_{l=1}^{\mathrm{M_T}} \mathbf{p}_{(l)}^H \mathbf{p}_{(l)} \odot \mathbf{e}_{(l)}^H \mathbf{e}_{(l)}.$$

Defining

$$\mathbf{R}_l = \mathbf{D}_{\mathbf{p}_{(l)}}^H \mathbf{R}_{\mathbb{H}}^T \mathbf{D}_{\mathbf{p}_{(l)}}, \quad l = 1, \dots, \mathrm{M_T} \tag{72}$$

and using $\mathbf{R}_l \odot \mathbf{e}_{(l)}^H \mathbf{e}_{(l)} = \mathbf{D}_{\mathbf{e}_{(l)}}^H \mathbf{R}_l \mathbf{D}_{\mathbf{e}_{(l)}}$ $(l = 1, \dots, \mathrm{M_T})$, the effective codeword difference matrix is given by

$$\mathbf{R}_{\mathbb{H}}^T \odot \tilde{\mathbf{E}}^H \tilde{\mathbf{E}} = \sum_{l=1}^{\mathrm{M_T}} \mathbf{D}_{\mathbf{e}_{(l)}}^H \mathbf{R}_l \mathbf{D}_{\mathbf{e}_{(l)}}. \tag{73}$$

Consequently, the code design criterion in Theorem 1 specializes to

$$\Xi_{\mathrm{m}}^{\rho \mathrm{M_T}}(\mathsf{SNR}) = \min_{\substack{\mathbf{E} = \mathbf{X} - \mathbf{X}', \mathbf{X} \neq \mathbf{X}' \\ \mathbf{X}, \mathbf{X}' \in \mathcal{C}_r(\mathsf{SNR})}} \prod_{k=1}^{\mathrm{m}} \lambda_k \left( \sum_{l=1}^{\mathrm{M_T}} \mathbf{D}_{\mathbf{e}_{(l)}}^H \mathbf{R}_l \mathbf{D}_{\mathbf{e}_{(l)}} \right) \dot{\geq} \mathsf{SNR}^{-(r-\epsilon)} \tag{74}$$

for some $\epsilon > 0$ constant w.r.t. SNR and $r$. We shall next formalize our main result in the context of code design for selective-fading MIMO channels.

*Theorem 2:* Consider a family of codes $\mathcal{C}_r$, $r \in [0, \mathrm{m}]$, of block length $N \geq \rho \mathrm{M_T}$. Let the transmit signal corresponding to antenna $l$, for $l = 1, \dots, \mathrm{M_T}$, be given by $\tilde{\mathbf{x}} = \mathbf{p}_{(l)} \odot \mathbf{x}$, where $\mathbf{x} = [x_0 \cdots x_{N-1}]$ is a codeword in $\mathcal{C}_r(\mathsf{SNR})$ and $\mathbf{p}_{(l)}$ is the $l$th row of the precoding matrix $\mathbf{P}$ $(\mathrm{M_T} \times N)$. If, for some $\epsilon > 0$ constant w.r.t. SNR and $r$, $\mathcal{C}_r$ satisfies

$$\min_{\substack{\mathbf{e} = \mathbf{x} - \mathbf{x}', \mathbf{x} \neq \mathbf{x}' \\ \mathbf{x}, \mathbf{x}' \in \mathcal{C}_r(\mathsf{SNR})}} \prod_{n=0}^{\mathrm{m}-1} |e_{\pi(n)}|^2 \dot{\geq} \mathsf{SNR}^{-(r-\epsilon)} \tag{75}$$

where $\pi$ is the (SNR-dependent) permutation that sorts the entries of $\mathbf{e}$ in ascending order for every SNR level[5], and $\mathbf{P}$ is such that

$$\mathrm{rank}\left( \mathbf{R}_{\mathbb{H}}^T \odot \mathbf{P}^H \mathbf{P} \right) = \rho \mathrm{M_T} \tag{76}$$

---

[5]Recall that the entries of $\mathbf{e}$ depend on SNR.

then the pair of inner and outer codes $(\mathbf{P}, \mathcal{C}_r)$ satisfies the code design criterion (45) in Theorem 1.

*Proof:* We start by noting that since the same $1 \times N$ codeword $\mathbf{x}$ is transmitted over all antennas, we have $\mathbf{e}_{(l)} = \mathbf{e}$, for all $l = 1, \ldots, \mathrm{M_T}$, which, upon inserting into (73), yields

$$\mathbf{R}_{\mathbb{H}}^T \odot \tilde{\mathbf{E}}^H \tilde{\mathbf{E}} = \sum_{l=1}^{\mathrm{M_T}} \mathbf{D}_{\mathbf{e}_{(l)}}^H \mathbf{R}_l \mathbf{D}_{\mathbf{e}_{(l)}} = \mathbf{D}_{\mathbf{e}}^H \left( \mathbf{R}_{\mathbb{H}}^T \odot \mathbf{P}^H \mathbf{P} \right) \mathbf{D}_{\mathbf{e}}. \qquad (77)$$

Condition (76) implies that exactly $\rho \mathrm{M_T}$ eigenvalues of $\mathbf{R}_{\mathbb{H}}^T \odot \mathbf{P}^H \mathbf{P}$ are nonzero (recall that $N \geq \rho \mathrm{M_T}$ so that $\mathrm{rank}\left( \mathbf{R}_{\mathbb{H}}^T \odot \mathbf{P}^H \mathbf{P} \right) \leq \min(N, \rho \mathrm{M_T}) = \rho \mathrm{M_T}$ is not limited by the block length $N$). With the eigenvalue decomposition $\mathbf{R}_{\mathbb{H}}^T \odot \mathbf{P}^H \mathbf{P} = \mathbf{V} \mathbf{\Sigma} \mathbf{V}^H$, where $\mathbf{\Sigma} = \mathrm{diag}\left\{ \tilde{\mathbf{\Sigma}}, 0, \ldots, 0 \right\}$, $\tilde{\mathbf{\Sigma}} = \mathrm{diag}\{\sigma_0, \ldots, \sigma_{\rho \mathrm{M_T}-1}\}$ and the nonzero eigenvalues $\sigma_i$ sorted in ascending order, we get $\mathbf{R}_{\mathbb{H}}^T \odot \tilde{\mathbf{E}}^H \tilde{\mathbf{E}} = \mathbf{D}_{\mathbf{e}}^H \mathbf{V} \mathbf{\Sigma} \mathbf{V}^H \mathbf{D}_{\mathbf{e}}$. Using the fact that $\lambda_n(\mathbf{M} \mathbf{M}^H) = \lambda_n(\mathbf{M}^H \mathbf{M})$, $\forall n$, for a square matrix $\mathbf{M}$, we obtain

$$\lambda_n(\mathbf{R}_{\mathbb{H}}^T \odot \tilde{\mathbf{E}}^H \tilde{\mathbf{E}}) = \lambda_n(\mathbf{\Sigma}^{1/2} \underbrace{\mathbf{V}^H \mathbf{D}_{\mathbf{e}} \mathbf{D}_{\mathbf{e}}^H \mathbf{V}}_{\triangleq \mathbf{B}} \mathbf{\Sigma}^{1/2})$$

$$= \lambda_n(\tilde{\mathbf{\Sigma}}^{1/2} \tilde{\mathbf{B}} \tilde{\mathbf{\Sigma}}^{1/2}) \qquad (78)$$

$$\geq \sigma_0 \, \lambda_n(\tilde{\mathbf{B}}) \qquad (79)$$

for the nonzero eigenvalues of $\mathbf{R}_{\mathbb{H}}^T \odot \tilde{\mathbf{E}}^H \tilde{\mathbf{E}}$, i.e., for $n = 0, \ldots, \rho \mathrm{M_T} - 1$. Here, $\tilde{\mathbf{B}} = \mathbf{B}([1 : \rho \mathrm{M_T}], [1 : \rho \mathrm{M_T}])$ and (79) follows by applying Ostrowski's Theorem [18, Theorem 4.5.9]. Since $\mathbf{B}$ is Hermitian and $\tilde{\mathbf{B}}$ is its principal submatrix obtained by deleting the $N - \rho \mathrm{M_T}$ last rows and the corresponding columns in $\mathbf{B}$, we can invoke [18, Theorem 4.3.15] to conclude that

$$\lambda_k(\tilde{\mathbf{B}}) \geq \lambda_k(\mathbf{B}) = |e_{\pi(k)}|^2, \; k = 0, \ldots, \rho \mathrm{M_T} - 1 \qquad (80)$$

where $\pi$ is the (SNR-dependent) permutation that sorts the entries of $\mathbf{e}$ in ascending order for every SNR value. Next, combining (79) with (80), we find that the nonzero[6] eigenvalues of $\mathbf{R}_{\mathbb{H}}^T \odot \tilde{\mathbf{E}}^H \tilde{\mathbf{E}}$ satisfy

$$\lambda_k(\mathbf{R}_{\mathbb{H}}^T \odot \tilde{\mathbf{E}}^H \tilde{\mathbf{E}}) \geq \sigma_0 \, |e_{\pi(k)}|^2, \; k = 0, \ldots, \rho \mathrm{M_T} - 1. \qquad (81)$$

---

[6]Recall that "nonzero eigenvalue" refers to an eigenvalue that is not identically equal to zero for all SNR values.

By (75), we can therefore conclude that

$$\Xi_{\mathrm{m}}^{\rho_{\mathrm{M_T}}}(\mathsf{SNR}) = \min_{\substack{\mathbf{E}=\mathbf{X}-\mathbf{X}', \mathbf{X}\neq\mathbf{X}' \\ \mathbf{X},\mathbf{X}'\in\mathcal{C}_r(\mathsf{SNR})}} \prod_{k=0}^{\mathrm{m}-1} \lambda_k(\mathbf{R}_{\widetilde{\mathbb{H}}}^T \odot \tilde{\mathbf{E}}^H \tilde{\mathbf{E}})$$

$$\geq (\sigma_0)^{\mathrm{m}} \min_{\substack{\mathbf{e}=\mathbf{x}-\mathbf{x}', \mathbf{x}\neq\mathbf{x}' \\ \mathbf{x},\mathbf{x}'\in\mathcal{C}_r(\mathsf{SNR})}} \prod_{n=0}^{\mathrm{m}-1} |e_{\pi(n)}|^2$$

$$\dot{\geq} \mathsf{SNR}^{-(r-\epsilon)}.$$

∎

The precoder $\mathbf{P}$ effectively decorrelates the channel into its independent diversity branches; the resulting design criterion for the outer family of codes (75) is satisfied by the QAM-based permutation codes proposed in [8] in the context of parallel channels. To see this, we start by recalling that the problem addressed in [8, Sec.V.B] is the construction of space-only codes, i.e., $N = 1$, that are approximately universal over a parallel channel with $L$ independent flat-fading subchannels. The code construction presented in [8] is based on permutations of QAM constellations. In order to sustain a rate of $R(\mathsf{SNR})$ over the parallel channel, each subchannel has as input alphabet a QAM constellation $\mathcal{A}(\mathsf{SNR})$ with $2^{R(\mathsf{SNR})}$ points. A permutation code across the $L$ subchannels can be represented as

$$\Pi(\mathsf{SNR}) = \Big\{ \mathbf{x} = [\pi_1(q) \ \ldots \ \pi_L(q)], q \in \mathcal{A}(\mathsf{SNR}) \Big\} \tag{82}$$

where $\mathcal{A}$ is the family of QAM constellations defined in (29) and the $\pi_l, l = 1, \ldots, L$, are permutations of the constellation elements in $\mathcal{A}(\mathsf{SNR})$. A remarkable result given in [8, Theorem 5.2] says that there exist permutations $\pi_l, l = 1, \ldots, L$, so that $\Pi$ in (82) constitutes an approximately universal code for the parallel channel. By [8, Theorem 5.1], such a family of codes $\Pi$ satisfies the following condition. Let $\mathbf{x}$ denote a codeword in $\Pi(\mathsf{SNR})$ as defined in (82), and denote the corresponding codeword difference vectors by $\mathbf{e} = \mathbf{x} - \mathbf{x}'$, $\mathbf{x} \neq \mathbf{x}'$, $\mathbf{x}, \mathbf{x}' \in \Pi(\mathsf{SNR})$. Then, the approximately universal family of codes $\Pi$ satisfies [8, Eq. (24)], i.e.,

$$|\mathbf{e}(1)|^2 \cdots |\mathbf{e}(L)|^2 \dot{\geq} \frac{1}{2^{R(\mathsf{SNR})-\epsilon \log \mathsf{SNR}}} = \mathsf{SNR}^{-(r-\epsilon)} \tag{83}$$

for all $\mathbf{e} \neq \mathbf{0}$ arising from $\Pi(\mathsf{SNR})$ and some $\epsilon > 0$ that is constant w.r.t. SNR and $r$.

Mapping the spatial dimension in (83) to time-frequency slots and setting $L = N$, it follows from [8, Th. 5.2] and (83) that there exist families of permutation codes $\Pi$ as given in (82) (now $\pi_n(q), n = 0, \ldots, N-1$, denotes the symbol transmitted in time-frequency slot $n$) that satisfy

$$|\mathbf{e}(1)|^2 \cdots |\mathbf{e}(N)|^2 \dot{\geq} \mathsf{SNR}^{-(r-\epsilon)} \tag{84}$$

for all $\mathbf{e} \neq \mathbf{0}$ arising from $\Pi(\mathsf{SNR})$ and some $\epsilon > 0$ constant w.r.t. SNR and $r$. Due to the power constraint (28) on the codewords of $\mathcal{C}_r$, we necessarily have $|\mathbf{e}(n)|^2 \stackrel{.}{\leq} 1$ for all $n$ so that (75) is satisfied. We can therefore conclude that the design criterion in Theorem 2 for the family of codes $\mathcal{C}_r$ can be satisfied using the QAM-based permutation codes proposed in [8]. We emphasize, however, that here coding is performed over time and frequency as opposed to [8] where coding is performed across parallel channels.

## VI. Precoder design

It remains to show that, given $\mathbf{R}_{\mathbb{H}}$, we can find a precoder $\mathbf{P}$ such that

$$\mathrm{rank}\big(\mathbf{R}_{\mathbb{H}}^T \odot \mathbf{P}^H \mathbf{P}\big) = \rho \mathrm{M_T}. \tag{85}$$

Using the eigenvalue decomposition $\mathbf{R}_{\mathbb{H}} = \sum_{n=0}^{\rho-1} \lambda_n \mathbf{u}_n \mathbf{u}_n^H$, we note that

$$\begin{aligned}
\mathbf{R}_{\mathbb{H}}^T \odot \mathbf{P}^H \mathbf{P} &= \left( \sum_{n=0}^{\rho-1} \lambda_n\, \mathbf{u}_n^* \mathbf{u}_n^T \right) \odot \left( \sum_{l=1}^{\mathrm{M_T}} \mathbf{p}_{(l)}^H \mathbf{p}_{(l)} \right) \\
&= \sum_{n=0}^{\rho-1} \sum_{l=1}^{\mathrm{M_T}} \lambda_n \underbrace{\mathbf{D}_{\mathbf{p}_{(l)}}^H \mathbf{u}_n^*}_{\boldsymbol{\alpha}_{n,l}} \underbrace{\mathbf{u}_n^T \mathbf{D}_{\mathbf{p}_{(l)}}}_{\boldsymbol{\alpha}_{n,l}^H}. 
\end{aligned} \tag{86}$$

The task of designing a precoder that satisfies (85) amounts to finding $\mathbf{p}_{(l)}$, $l = 1, \ldots, \mathrm{M_T}$, such that the corresponding $\boldsymbol{\alpha}_{n,l}$ are linearly independent. Enforcing structure in $\mathbf{R}_{\mathbb{H}}$ allows to get more specific about how to design the precoder. This can be illustrated as follows.

*Example:* Consider the case of cyclic ISI channels (e.g., OFDM modulation) with $\mathrm{M_T} = 2$, $L = 2$, and $N = 4$. Using (56) the corresponding covariance matrix is obtained as $\mathbf{R}_{\mathbb{H}} = \lambda_0 \boldsymbol{\psi}_0 \boldsymbol{\psi}_0^H + \lambda_1 \boldsymbol{\psi}_1 \boldsymbol{\psi}_1^H$, where the eigenvectors of $\mathbf{R}_{\mathbb{H}}$ are simply columns of the FFT matrix $\boldsymbol{\Psi} = [\boldsymbol{\psi}_0\ \boldsymbol{\psi}_1\ \boldsymbol{\psi}_2\ \boldsymbol{\psi}_3]$, i.e., $\mathbf{u}_n = \boldsymbol{\psi}_n$, $n = 0, \ldots, 3$. One possibility to obtain a set of linearly independent vectors $\boldsymbol{\alpha}_{n,l}$ in (86) is to set

$$\mathbf{p}_{(l)} = \boldsymbol{\psi}_{(l-1)L}^T, \quad l = 1, 2. \tag{87}$$

More concretely, invoking

$$\mathbf{D}_{\boldsymbol{\psi}_m}^H \boldsymbol{\psi}_n^* = \boldsymbol{\psi}_{(n+m)\bmod N}^*$$

the precoder defined through (87) results in

$$\begin{aligned}
\mathbf{R}_{\mathbb{H}}^T \odot \mathbf{P}^H \mathbf{P} &= \mathbf{D}_{\boldsymbol{\psi}_0}^H \big( \lambda_0 \boldsymbol{\psi}_0^* \boldsymbol{\psi}_0^T + \lambda_1 \boldsymbol{\psi}_1^* \boldsymbol{\psi}_1^T \big)\, \mathbf{D}_{\boldsymbol{\psi}_0} \\
&\quad + \mathbf{D}_{\boldsymbol{\psi}_2}^H \big( \lambda_0 \boldsymbol{\psi}_0^* \boldsymbol{\psi}_0^T + \lambda_1 \boldsymbol{\psi}_1^* \boldsymbol{\psi}_1^T \big)\, \mathbf{D}_{\boldsymbol{\psi}_2}
\end{aligned}$$

$$= \lambda_0 \boldsymbol{\psi}_0^* \boldsymbol{\psi}_0^T + \lambda_1 \boldsymbol{\psi}_1^* \boldsymbol{\psi}_1^T$$

$$+ \lambda_0 \boldsymbol{\psi}_2^* \boldsymbol{\psi}_2^T + \lambda_1 \boldsymbol{\psi}_3^* \boldsymbol{\psi}_3^T$$

$$= \boldsymbol{\Psi}^* \operatorname{diag}\{\lambda_0, \lambda_1, \lambda_0, \lambda_1\} \boldsymbol{\Psi}^T$$

which is clearly a full-rank matrix. Note that this precoder simply amounts to performing (cylic) delay diversity.

We next consider general time-frequency selective channels where the corresponding covariance matrix $\mathbf{R}_{\mathbb{H}}$—as a consequence of the stationarity of $L_{\mathbb{H}}(t, f)$ in $t$ and $f$—is two-level Toeplitz[7]. In this case, it seems difficult to devise a general analytic procedure for constructing $\mathbf{P}$ for a given $\mathbf{R}_{\mathbb{H}}$ such that (85) is satisfied. We can, however, exploit the asymptotic equivalence of two-level Toeplitz and two-level circulant matrices to satisfy (85) asymptotically in the block length $N$. In particular, we will need the following result.

*Theorem 3 (Asymptotic Eigenvalue Distribution [22]–[24]):* The distribution of the eigenvalues of $\mathbf{R}_{\mathbb{H}}$ for $M, K \to \infty$, where $M$ and $K$ are related to the block length $N$ by the mapping (11), is given by

$$S(\xi, \mu) = \sum_{m=-\infty}^{\infty} \sum_{k=-\infty}^{\infty} R_{\mathbb{H}}(mT, kF) \, e^{-j2\pi(\mu m - \xi k)}$$

$$= \frac{1}{TF} \sum_{i=-\infty}^{\infty} \sum_{j=-\infty}^{\infty} C_{\mathbb{H}}\left(\frac{\xi + i}{F}, \frac{\mu + j}{T}\right), \quad 0 \le \mu, \xi < 1.$$

In what follows, we design the precoder $\mathbf{P}$ based on a (two-level) circulant approximation $\mathbf{C}_{\mathbb{H}}$ of the (two-level) Toeplitz covariance matrix $\mathbf{R}_{\mathbb{H}}$. Specifically, we take the matrix $\mathbf{C}_{\mathbb{H}}$ such that its eigenvalues are uniformly-spaced samples of the asymptotic eigenvalue distribution of $\mathbf{R}_{\mathbb{H}}$ given by $S(\xi, \mu)$. This implies that $\mathbf{C}_{\mathbb{H}}$ and $\mathbf{R}_{\mathbb{H}}$ are asymptotically (in block length $N$) equivalent [22, Lemma 11], [23, Lemma 1] and that their eigenvalues are asymptotically equally distributed[8] [22, Theorem 9], [23, Theorem 1]. In cases where the signal model is (two-level) circulant [14], [16], this approach gives exact results for any block length $N$ because $\mathbf{R}_{\mathbb{H}}$ is (two-level) circulant for any $K$ and $M$. For general (two-level) Toeplitz covariance matrices $\mathbf{R}_{\mathbb{H}}$, this approach is meaningful because the asymptotic equivalence of $\mathbf{C}_{\mathbb{H}}$ and $\mathbf{R}_{\mathbb{H}}$ implies asymptotic equivalence of $\mathbf{C}_{\mathbb{H}}^T \odot \mathbf{P}^H \mathbf{P}$ and $\mathbf{R}_{\mathbb{H}}^T \odot \mathbf{P}^H \mathbf{P}$.

---

[7]A two-level Toeplitz matrix is a block Toeplitz matrix with Toeplitz blocks. Similarly, a two-level circulant matrix is a block circulant matrix with circulant blocks.

[8]The interested reader is referred to [22, Theorem 4] (respectively, [23, Theorem 2]) for a formal definition of the concept of asymptotically equally distributed one-dimensional (or two-dimensional) sequences.

We start by defining the (two-level) circulant matrix

$$\mathbf{C}_{\mathbb{H}} = \mathbf{F} \boldsymbol{\Lambda} \mathbf{F}^H$$

where $\mathbf{F} = \boldsymbol{\Psi} \otimes \boldsymbol{\Phi}$, with $\boldsymbol{\Psi}$ and $\boldsymbol{\Phi}$ denoting the $M \times M$ and $K \times K$ FFT matrices, respectively, and $\boldsymbol{\Lambda} = \mathrm{diag}\{\lambda_n(\mathbf{C}_{\mathbb{H}})\}_{n=0}^{N-1}$, with

$$\lambda_n(\mathbf{C}_{\mathbb{H}}) \triangleq S\left(\frac{k}{K}, \frac{m}{M}\right), \quad m = 0, \ldots, M-1, \ k = 0, \ldots, K-1 \tag{88}$$

where we have used the mapping $n = \mathcal{M}(m, k)$ defined in (11). Because the scattering function is assumed to be compactly supported in the rectangle $[0, \tau_0] \times [0, \nu_0]$, $S(\xi, \mu)$ is also compactly supported, and hence the nonzero eigenvalues of $\mathbf{R}_{\mathbb{H}}$ in (88) are indexed by

$$(m, k) \in \{0, \ldots, v-1\} \times \{0, \ldots, t-1\} \tag{89}$$

where

$$v \triangleq \lfloor \nu_0 T M \rfloor \quad \text{and} \quad t \triangleq \lfloor \tau_0 F K \rfloor. \tag{90}$$

Next, we propose a precoder tailored to $\mathbf{C}_{\mathbb{H}}$ that achieves $\mathrm{rank}(\mathbf{C}_{\mathbb{H}}^T \odot \mathbf{P}^H \mathbf{P}) = \rho \mathrm{M_T}$. The main idea underlying this construction is to design $\mathbf{P}$ such that the precoder effectively induces time-frequency shifts with the shifts chosen appropriately.

*Proposition 3:* Consider the $N \times N$ matrix $\mathbf{C}_{\mathbb{H}} = \mathbf{F} \boldsymbol{\Lambda} \mathbf{F}^H$, where $\mathbf{F} = \boldsymbol{\Psi} \otimes \boldsymbol{\Phi}$ ($\boldsymbol{\Psi}$, $\boldsymbol{\Phi}$ are the $M \times M$ and $K \times K$ FFT matrices, respectively) and $\boldsymbol{\Lambda}$ has $\rho = vt$ nonzero diagonal elements. If $N \geq \rho \mathrm{M_T}$ and $\mathbf{P}$ satisfies

$$\mathbf{p}_{(l)}^T = \boldsymbol{\psi}_{p_l v} \otimes \boldsymbol{\phi}_{q_l t}, \text{ for } l = 1, \ldots, \mathrm{M_T} \tag{91}$$

where $\boldsymbol{\psi}_m$ and $\boldsymbol{\phi}_k$ are, respectively, the $m$th and $k$th columns of $\boldsymbol{\Psi}$ and $\boldsymbol{\Phi}$, and

$$(p_l, q_l) \in \left\{0, \ldots, \left\lfloor \frac{1}{\nu_0 T} \right\rfloor - 1\right\} \times \left\{0, \ldots, \left\lfloor \frac{1}{\tau_0 F} \right\rfloor - 1\right\}, (p_l, q_l) \neq (p_{l'}, q_{l'}) \text{ for } l \neq l', \tag{92}$$

then $\mathrm{rank}(\mathbf{C}_{\mathbb{H}}^T \odot \mathbf{P}^H \mathbf{P}) = \rho \mathrm{M_T}$.

*Proof:* We start by noting that $\mathbf{C}_{\mathbb{H}}^T \odot \mathbf{P}^H \mathbf{P}$ can be written as

$$\mathbf{C}_{\mathbb{H}}^T \odot \mathbf{P}^H \mathbf{P} = \sum_{l=1}^{\mathrm{M_T}} \underbrace{\mathbf{D}_{\mathbf{p}_{(l)}^H} \mathbf{C}_{\mathbb{H}}^T \mathbf{D}_{\mathbf{p}_{(l)}}}_{\triangleq \, \mathbf{C}_l}. \tag{93}$$

Next, consider the following similarity transformation

$$\mathbf{F}^T \mathbf{C}_l \mathbf{F}^* = \mathbf{F}^T \mathbf{D}_{\mathbf{p}_{(l)}^H} \mathbf{F}^* \boldsymbol{\Lambda} \mathbf{F}^T \mathbf{D}_{\mathbf{p}_{(l)}} \mathbf{F}^* \tag{94}$$

where we have used $\mathbf{C}_{\mathbb{H}} = \mathbf{F}\boldsymbol{\Lambda}\mathbf{F}^H$. With (91) and $\mathbf{F} = \boldsymbol{\Psi} \otimes \boldsymbol{\Phi}$, we get

$$\mathbf{F}^T \mathbf{D}_{\mathbf{p}_{(l)}^H} \mathbf{F}^* = \left( \boldsymbol{\Psi}^T \mathbf{D}_{\boldsymbol{\psi}_{p_l v}^*} \boldsymbol{\Psi}^* \right) \otimes \left( \boldsymbol{\Phi}^T \mathbf{D}_{\boldsymbol{\phi}_{q_l t}^*} \boldsymbol{\Phi}^* \right)$$

$$= \boldsymbol{\Pi}^{p_l v} \otimes \boldsymbol{\Pi}^{q_l t} \tag{95}$$

where $\boldsymbol{\Pi} = [\boldsymbol{\pi}_1 \cdots \boldsymbol{\pi}_{N-1} \, \boldsymbol{\pi}_0]$, with $\boldsymbol{\pi}_n = [0 \cdots 0 \, 1 \, 0 \cdots 0]^T$ containing a $1$ in its $n$th position, is the circulant permutation matrix. Using (95) in (94), we obtain

$$\mathbf{F}^T \mathbf{C}_l \mathbf{F}^* = \left( \boldsymbol{\Pi}^{p_l v} \otimes \boldsymbol{\Pi}^{q_l t} \right) \boldsymbol{\Lambda} \left( \boldsymbol{\Pi}^{p_l v} \otimes \boldsymbol{\Pi}^{q_l t} \right)^T \tag{96}$$

and consequently

$$\mathbf{F}^T \left( \mathbf{C}_{\mathbb{H}}^T \odot \mathbf{P}^H \mathbf{P} \right) \mathbf{F}^* = \sum_{l=1}^{M_T} \left( \boldsymbol{\Pi}^{p_l v} \otimes \boldsymbol{\Pi}^{q_l t} \right) \boldsymbol{\Lambda} \left( \boldsymbol{\Pi}^{p_l v} \otimes \boldsymbol{\Pi}^{q_l t} \right)^T. \tag{97}$$

Since $\left( \boldsymbol{\Pi}^k \otimes \boldsymbol{\Pi}^l \right) \boldsymbol{\Lambda} \left( \boldsymbol{\Pi}^k \otimes \boldsymbol{\Pi}^l \right)^T$ simply permutes the entries of $\boldsymbol{\Lambda}$ along the main diagonal, the rank of $\mathbf{C}_{\mathbb{H}}^T \odot \mathbf{P}^H \mathbf{P}$ is trivially bounded above by $\rho M_T$. To achieve this maximum rank, we need to ensure that the different shifts in (97) distribute the $\rho$ eigenvalues of $\mathbf{C}_{\mathbb{H}}$ into mutually orthogonal subspaces. This can be accomplished as follows. With (89) and (96), we find that the indices $(m, k)$ corresponding to the nonzero eigenvalues of $\mathbf{C}_l$ are given by the set

$$\mathcal{I}_l = \{ p_l v, \ldots, (p_l + 1)v - 1 \} \times \{ q_l t, \ldots, (q_l + 1)t - 1 \}$$

that is, the nonzero eigenvalues of $\mathbf{C}_l$ are obtained by cyclically shifting the eigenvalues of $\mathbf{C}_{\mathbb{H}}$ by $p_l v$ positions along index $m$ and $q_l t$ positions along index $k$. The condition in (92) guarantees that $\mathcal{I}_l \cap \mathcal{I}_{l'} = \emptyset$ for $l \neq l'$, which together with $\rho = vt$ in turn ensures that $\operatorname{rank}\left( \mathbf{C}_{\mathbb{H}}^T \odot \mathbf{P}^H \mathbf{P} \right) = \rho M_T$.

∎

We finally note that the precoder described in Proposition 3 is a generalization of well-known transmit diversity techniques that convert spatial diversity into time or frequency diversity [25]–[27]. This can be seen as follows. From (91), we note that the precoder $\mathbf{P}$ amounts to multiplying the signal transmitted from the $l$th antenna by

$$\mathbf{p}_{(l)}(n) = \exp\left( -j2\pi \left( p_l v \frac{m}{M} + q_l t \frac{k}{K} \right) \right), \quad \text{for } n = 0, \ldots, N-1 \tag{98}$$

where the pair $(m, k)$ is related to the slot index $n$ by $\mathcal{M}(m, k) = n$. For $K = 1$ (and hence $k = 0$, and $N = M$ in (98)), the index $n = m$ runs over time, resulting in

$$\mathbf{p}_{(l)}(n) = \exp\left( -j \frac{2\pi n}{M} p_l v \right), \quad \text{for } n = 0, \ldots, M-1 \tag{99}$$

which shows that the precoder simply introduces a frequency offset across transmit antennas—a technique known as phase rolling [27]–[32]. On the other hand, for $M = 1$ (and hence $m = 0$, and $N = K$ in (98)), the index $n = k$ runs over frequency and we obtain

$$\mathbf{p}_{(l)}(n) = \exp\left(-j\frac{2\pi n}{K}q_l t\right), \quad \text{for } n = 0, \ldots, K-1 \tag{100}$$

which shows that the precoder induces a time offset, i.e., a delay, across transmit antennas and hence corresponds to delay diversity as proposed in [25], [26], [31], [32]. In the case of general $M$ and $K$, the precoder in (98) induces time and frequency shifts. While delay diversity and phase rolling are well-known and easy-to-implement transmit diversity techniques for MISO systems that have been shown to have the potential of realizing full diversity gain for $r = 0$, it is surprising to see that they result in DM tradeoff optimality (when combined with proper outer codes) for multiplexing rates greater than zero.

## VII. CONCLUSION

Analyzing the high-SNR outage behavior of the Jensen channel instead of the original channel was found to be an effective tool for establishing the optimal DM tradeoff in general selective-fading MIMO channels. Our achievability proof reveals a code design criterion for DM tradeoff optimality based on which it is shown that the code design problem can be solved in a systematic fashion by combining a precoder adapted to the channel statistics with an outer code that is DM tradeoff optimal for parallel fading channels. The main result of the paper is supported by an appealing geometric argument, first provided in the flat-fading case in [1]. Finally, we note that the concepts introduced in this paper can be extended to multiple-access selective-fading MIMO channels [33] and to the analysis of the DM tradeoff properties of relay channels [34].

## APPENDIX I
## PROOF OF THEOREM 1

We start by deriving an upper bound on the average (w.r.t. the random channel) pairwise error probability (PEP). Assuming that $\mathbf{X} = [\mathbf{x}_0 \cdots \mathbf{x}_{N-1}]$ was transmitted, the probability of the ML decoder mistakenly deciding in favor of codeword $\mathbf{X}' = [\mathbf{x}'_0 \cdots \mathbf{x}'_{N-1}]$ can be upper-bounded in terms of the codeword difference matrix $\mathbf{E} = [\mathbf{e}_0 \cdots \mathbf{e}_{N-1}]$ with $\mathbf{e}_n = \mathbf{x}_n - \mathbf{x}'_n$ as

$$\begin{aligned}
\mathbb{P}(\mathbf{X} \to \mathbf{X}') &\leq \mathbb{E}_{\mathbf{H}}\left\{\exp\left(-\frac{\mathsf{SNR}}{4\mathrm{M_T}}\sum_{n=0}^{N-1}||\mathbf{H}_n\mathbf{e}_n||^2\right)\right\} \\
&= \mathbb{E}_{\mathbf{H}}\left\{\exp\left(-\frac{\mathsf{SNR}}{4\mathrm{M_T}}\mathrm{Tr}\left(\mathbf{H}_w\mathbf{\Upsilon}\mathbf{\Upsilon}^H\mathbf{H}_w^H\right)\right)\right\}
\end{aligned} \tag{101}$$

where (101) is the Chernoff bound on the PEP, $\mathbf{H}_w$ denotes an $\mathrm{M_R} \times \mathrm{M_T} N$ i.i.d. $\mathcal{CN}(0, 1)$ matrix, and we have introduced the matrix

$$\mathbf{\Upsilon} = (\mathbf{R}_{\boxplus}^{T/2} \otimes \mathbf{I}_{\mathrm{M_T}}) \operatorname{diag}\{\mathbf{e}_n\}_{n=0}^{N-1}. \tag{102}$$

Noting that

$$\mathbf{\Upsilon}^H \mathbf{\Upsilon} = \mathbf{R}_{\boxplus}^T \odot \mathbf{E}^H \mathbf{E} \tag{103}$$

and using the fact that the nonzero[9] eigenvalues of $\mathbf{\Upsilon}^H \mathbf{\Upsilon}$ equal the nonzero eigenvalues of $\mathbf{\Upsilon} \mathbf{\Upsilon}^H$ for every SNR, it follows, by assumption, that $\mathbf{\Upsilon} \mathbf{\Upsilon}^H$ has $\rho \mathrm{M_T}$ nonzero eigenvalues denoted as $\lambda_1(\mathsf{SNR}) \leq \lambda_2(\mathsf{SNR}) \leq \cdots \leq \lambda_{\rho \mathrm{M_T}}(\mathsf{SNR})$ (see Sec. IV-A). Then, performing an eigenvalue decomposition according to $\mathbf{\Upsilon} \mathbf{\Upsilon}^H = \mathbf{U} \mathbf{\Lambda} \mathbf{U}^H$, where the $N M_R \times N M_T$ matrix $\mathbf{U}$ is unitary and $\mathbf{\Lambda} = \operatorname{diag}\{\bar{\mathbf{\Lambda}}, \mathbf{0}\}$ with $\bar{\mathbf{\Lambda}} = \operatorname{diag}\{\lambda_k(\mathsf{SNR})\}_{k=1}^{\rho \mathrm{M_T}}$, we have $\operatorname{Tr}\left(\mathbf{H}_w \mathbf{\Upsilon} \mathbf{\Upsilon}^H \mathbf{H}_w^H\right) \sim \operatorname{Tr}\left(\mathbf{H}_w \mathbf{\Lambda} \mathbf{H}_w^H\right)$. Hence, setting $\overline{\mathbf{H}}_w = \mathbf{H}_w([1{:}\mathrm{M_R}], [1{:}\rho \mathrm{M_T}])$, it follows that

$$\mathbb{P}(\mathbf{X} \to \mathbf{X}') \leq \mathbb{E}_{\mathbf{H}_w}\left\{\exp\left(-\frac{\mathsf{SNR}}{4\mathrm{M_T}} \operatorname{Tr}\left(\overline{\mathbf{H}}_w \bar{\mathbf{\Lambda}} \overline{\mathbf{H}}_w^H\right)\right)\right\}. \tag{104}$$

Next, we express the right-hand side (RHS) of (104) in terms of the Jensen channel $\mathcal{H} = \mathcal{H}_w(\mathbf{R}^{T/2} \otimes \mathbf{I_M})$, where $\mathbf{R} = \mathbf{R}_{\boxplus}$, if $\mathrm{M_R} \leq \mathrm{M_T}$, and $\mathbf{R} = \mathbf{R}_{\boxplus}^T$, if $\mathrm{M_R} > \mathrm{M_T}$, and $\mathcal{H}_w$ is defined in (32).

For $\mathrm{M_R} \leq \mathrm{M_T}$, we note that $\overline{\mathbf{H}}_w = \overline{\mathcal{H}}_w$, with $\overline{\mathcal{H}}_w = \mathcal{H}_w([1{:}\mathrm{M_R}], [1{:}\rho \mathrm{M_T}])$. Invoking Theorem 4 in Appendix II, we get

$$\operatorname{Tr}\left(\overline{\mathbf{H}}_w \bar{\mathbf{\Lambda}} \overline{\mathbf{H}}_w^H\right) \geq \sum_{k=1}^{\mathrm{M_R}} \lambda_k(\overline{\mathbf{H}}_w \overline{\mathbf{H}}_w^H) \, \lambda_{\mathrm{M_R}+1-k}(\mathsf{SNR})$$

$$= \sum_{k=1}^{\mathrm{M_R}} \lambda_k(\overline{\mathcal{H}}_w \overline{\mathcal{H}}_w^H) \, \lambda_{\mathrm{M_R}+1-k}(\mathsf{SNR}). \tag{105}$$

For $\mathrm{M_R} > \mathrm{M_T}$, we set $\bar{\mathbf{\Lambda}} = \operatorname{diag}\{\bar{\mathbf{\Lambda}}_n\}_{n=0}^{\rho-1}$, where $\bar{\mathbf{\Lambda}}_n = \operatorname{diag}\{\lambda_k\}_{k=n\mathrm{M_T}+1}^{(n+1)\mathrm{M_T}}$, to get

$$\operatorname{Tr}\left(\overline{\mathbf{H}}_w \bar{\mathbf{\Lambda}} \overline{\mathbf{H}}_w^H\right) = \sum_{n=0}^{\rho-1} \operatorname{Tr}\left(\mathbf{H}_{w,n} \bar{\mathbf{\Lambda}}_n \mathbf{H}_{w,n}^H\right) \tag{106}$$

where $\overline{\mathbf{H}}_w = [\mathbf{H}_{w,0} \; \cdots \; \mathbf{H}_{w,\rho-1}]$. Because the eigenvalue ordering implies $\bar{\mathbf{\Lambda}}_0 \preceq \bar{\mathbf{\Lambda}}_n$ for all $n \neq 0$, we can invoke [18, Observation 7.7.2, Corollary 7.7.4(b)] to write $\operatorname{Tr}\left(\mathbf{H}_{w,n} \bar{\mathbf{\Lambda}}_n \mathbf{H}_{w,n}^H\right) \geq \operatorname{Tr}\left(\mathbf{H}_{w,n} \bar{\mathbf{\Lambda}}_0 \mathbf{H}_{w,n}^H\right)$ for all $n \neq 0$. Now (106) can be rewritten as

$$\sum_{n=0}^{\rho-1} \operatorname{Tr}\left(\mathbf{H}_{w,n} \bar{\mathbf{\Lambda}}_n \mathbf{H}_{w,n}^H\right) \geq \sum_{n=0}^{\rho-1} \operatorname{Tr}\left(\mathbf{H}_{w,n} \bar{\mathbf{\Lambda}}_0 \mathbf{H}_{w,n}^H\right)$$

----

[9]We recall that "nonzero eigenvalue" refers to an eigenvalue that is not identically equal to zero for all SNR values.

$$= \sum_{n=0}^{\rho-1} \mathrm{Tr}\left(\bar{\boldsymbol{\Lambda}}_0^{1/2} \mathbf{H}_{w,n}^H \mathbf{H}_{w,n} \bar{\boldsymbol{\Lambda}}_0^{1/2}\right)$$

$$= \mathrm{Tr}\left(\bar{\boldsymbol{\Lambda}}_0^{1/2} \left(\sum_{n=0}^{\rho-1} \mathbf{H}_{w,n}^H \mathbf{H}_{w,n}\right) \bar{\boldsymbol{\Lambda}}_0^{1/2}\right)$$

$$= \mathrm{Tr}\left(\bar{\boldsymbol{\Lambda}}_0^{1/2} \overline{\boldsymbol{\mathcal{H}}}_w \overline{\boldsymbol{\mathcal{H}}}_w^H \bar{\boldsymbol{\Lambda}}_0^{1/2}\right) \tag{107}$$

$$\geq \sum_{k=1}^{\mathrm{M_T}} \lambda_k(\overline{\boldsymbol{\mathcal{H}}}_w \overline{\boldsymbol{\mathcal{H}}}_w^H)\, \lambda_{\mathrm{M_T}+1-k}(\mathsf{SNR}) \tag{108}$$

where we set $\overline{\boldsymbol{\mathcal{H}}}_w = \boldsymbol{\mathcal{H}}_w([1:\mathrm{M_T}],[1:\rho\mathrm{M_R}])$ with $\boldsymbol{\mathcal{H}}_w$ given by (32) to get (107), and (108) follows immediately upon applying Theorem 4 in Appendix II to (107). Combining (105) and (108), we have, for general $\mathrm{M_T}$ and $\mathrm{M_R}$, that

$$\mathrm{Tr}\left(\overline{\mathbf{H}}_w \boldsymbol{\Lambda} \overline{\mathbf{H}}_w^H\right) \geq \sum_{k=1}^{\mathrm{m}} \lambda_k(\overline{\boldsymbol{\mathcal{H}}}_w \overline{\boldsymbol{\mathcal{H}}}_w^H)\lambda_{\mathrm{m}+1-k}(\mathsf{SNR})$$

$$= \sum_{k=1}^{\mathrm{m}} \mathsf{SNR}^{-\alpha_k}\lambda_{\mathrm{m}+1-k}(\mathsf{SNR}) \tag{109}$$

where (109) follows from the definition in (36). Using (109) in (104), we obtain a PEP upper bound in terms of the singularity levels $\alpha_k$ $(k=1,\ldots,\mathrm{m})$ characterizing the Jensen outage event

$$\mathbb{P}(\mathbf{X} \to \mathbf{X}') \leq \mathbb{E}_{\boldsymbol{\alpha}}\left\{\exp\left(-\frac{1}{4\mathrm{M_T}} \sum_{k=1}^{\mathrm{m}} \mathsf{SNR}^{1-\alpha_k}\, \lambda_{\mathrm{m}+1-k}(\mathsf{SNR})\right)\right\}. \tag{110}$$

Next, consider a realization of the random vector $\boldsymbol{\alpha}$ and let $\mathcal{S}_{\boldsymbol{\alpha}} = \{k : \alpha_k \leq 1\}$. We have

$$\sum_{k=1}^{\mathrm{m}} \mathsf{SNR}^{1-\alpha_k}\, \lambda_{\mathrm{m}+1-k}(\mathsf{SNR}) \geq \sum_{k \in \mathcal{S}_{\boldsymbol{\alpha}}} \mathsf{SNR}^{1-\alpha_k}\, \lambda_{\mathrm{m}+1-k}(\mathsf{SNR})$$

$$\geq |\mathcal{S}_{\boldsymbol{\alpha}}| \left(\mathsf{SNR}^{\sum_{k=1}^{\mathrm{m}}[1-\alpha_k]^+} \prod_{k \in \mathcal{S}_{\boldsymbol{\alpha}}} \lambda_{\mathrm{m}+1-k}(\mathsf{SNR})\right)^{\frac{1}{|\mathcal{S}_{\boldsymbol{\alpha}}|}} \tag{111}$$

where we used the arithmetic-geometric mean inequality and

$$\sum_{k \in \mathcal{S}_{\boldsymbol{\alpha}}} (1-\alpha_k) = \sum_{k=1}^{\mathrm{m}} [1-\alpha_k]^+$$

is an immediate consequence of the definition of $\mathcal{S}_{\boldsymbol{\alpha}}$. Using (111) in (110), we obtain

$$\mathbb{P}(\mathbf{X} \to \mathbf{X}') \leq \mathbb{E}_{\boldsymbol{\alpha}}\left\{\exp\left(-\frac{|\mathcal{S}_{\boldsymbol{\alpha}}|}{4\mathrm{M_T}} \left(\mathsf{SNR}^{\sum_{k=1}^{\mathrm{m}}[1-\alpha_k]^+} \prod_{k \in \mathcal{S}_{\boldsymbol{\alpha}}} \lambda_{\mathrm{m}+1-k}(\mathsf{SNR})\right)^{\frac{1}{|\mathcal{S}_{\boldsymbol{\alpha}}|}}\right)\right\}. \tag{112}$$

The dependency of the PEP upper bound (112) on the singularity levels characterizing the Jensen outage event suggests to split up the error probability according to

$$P_e(\mathcal{C}_r) = \mathbb{P}(\text{error}, \boldsymbol{\alpha} \in \mathcal{J}_r) + \mathbb{P}(\text{error}, \boldsymbol{\alpha} \notin \mathcal{J}_r)$$

$$= \mathbb{P}(\mathcal{J}_r)\,\mathbb{P}(\text{error}|\boldsymbol{\alpha} \in \mathcal{J}_r) + \mathbb{P}(\bar{\mathcal{J}}_r)\,\mathbb{P}(\text{error}|\boldsymbol{\alpha} \notin \mathcal{J}_r)$$

$$\leq \mathbb{P}(\mathcal{J}_r) + \mathbb{P}(\bar{\mathcal{J}}_r)\,\mathbb{P}(\text{error}|\boldsymbol{\alpha} \notin \mathcal{J}_r)\,. \tag{113}$$

For any $\boldsymbol{\alpha} \notin \mathcal{J}_r$ with $r > 0$, we have, by definition, $\sum_{k=1}^{\mathrm{m}}[1 - \alpha_k]^+ \geq r$ and consequently $|\mathcal{S}_{\boldsymbol{\alpha}}| \geq 1$, which upon noting that $|\mathcal{C}_r(\mathsf{SNR})| = \mathsf{SNR}^{N_r}$, yields the following union bound based on the PEP in (112)

$$\mathbb{P}(\text{error}|\boldsymbol{\alpha} \notin \mathcal{J}_r) \leq \mathsf{SNR}^{N_r} \exp\left(-\frac{1}{4\mathrm{M_T}}\left(\mathsf{SNR}^r \prod_{k \in \mathcal{S}_{\boldsymbol{\alpha}}} \lambda_{\mathrm{m}+1-k}(\mathsf{SNR})\right)^{\frac{1}{\mathrm{m}}}\right) \tag{114}$$

where we used $|\mathcal{S}_{\boldsymbol{\alpha}}| \leq \mathrm{m}$. Next, we note that the code design criterion in (45) implies that $\prod_{k=1}^{\mathrm{m}} \lambda_k(\mathsf{SNR}) \dot{\geq} \mathsf{SNR}^{-(r-\epsilon)}$ for some $\epsilon > 0$ that is constant w.r.t. SNR and $r$. Recalling from (43) that $\lambda_k(\mathsf{SNR}) \dot{\leq} 1$ for all $k$, we necessarily have

$$\prod_{k \in \mathcal{S}_{\boldsymbol{\alpha}}} \lambda_{\mathrm{m}+1-k}(\mathsf{SNR}) \dot{\geq} \mathsf{SNR}^{-(r-\epsilon)} \tag{115}$$

for any $\mathcal{S}_{\boldsymbol{\alpha}} \subseteq \{1, \ldots, \mathrm{m}\}$. Using (115) in (114), we get

$$\mathbb{P}(\text{error}, \boldsymbol{\alpha} \notin \mathcal{J}_r) = \underbrace{\mathbb{P}(\bar{\mathcal{J}}_r)}_{\leq 1}\,\mathbb{P}(\text{error}|\boldsymbol{\alpha} \notin \mathcal{J}_r)$$

$$\dot{\leq} \mathsf{SNR}^{N_r} \exp\left(-\frac{\mathsf{SNR}^{\epsilon/\mathrm{m}}}{4\mathrm{M_T}}\right)\,. \tag{116}$$

In contrast to the Jensen outage probability which satisfies $\mathbb{P}(\mathcal{J}_r) \doteq \mathsf{SNR}^{-d_{\mathcal{J}}(r)}$, the quantity on the RHS of (116) decays exponentially in SNR for any $r > 0$. Hence, upon inserting (116) in (113), we obtain

$$P_e(\mathcal{C}_r) \dot{\leq} \mathbb{P}(\mathcal{J}_r) \tag{117}$$

for $r > 0$. Since $\mathbb{P}(\mathcal{J}_r) \leq \mathbb{P}(\mathcal{O}_r)$, it follows trivially that $\mathbb{P}(\mathcal{J}_r) \dot{\leq} \mathbb{P}(\mathcal{O}_r)$. In addition, for a specific family of codes $\mathcal{C}_r$, we have $\mathbb{P}(\mathcal{O}_r) \leq P_e(\mathcal{C}_r)$ and hence $\mathbb{P}(\mathcal{O}_r) \dot{\leq} P_e(\mathcal{C}_r)$. Putting the pieces together, thanks to (117), we obtain that for any $r > 0$

$$\mathbb{P}(\mathcal{O}_r) \dot{\leq} P_e(\mathcal{C}_r) \dot{\leq} \mathbb{P}(\mathcal{J}_r) \dot{\leq} \mathbb{P}(\mathcal{O}_r)$$

which implies that

$$P_e(\mathcal{C}_r) \doteq \mathbb{P}(\mathcal{J}_r) \doteq \mathbb{P}(\mathcal{O}_r)$$

and hence, by definition of $d_{\mathcal{J}}(r)$, we get

$$P_e(\mathcal{C}_r) \doteq \mathsf{SNR}^{-d_{\mathcal{J}}(r)}. \tag{118}$$

Finally, as (118) holds for any $r > 0$ arbitrarily close to zero, we can invoke the continuity of the piecewise linear function $d_{\mathcal{J}}(r)$ to conclude that (118) also holds in the limit $r \to 0$ [1, Proof of Lemma 5], hence establishing the desired result.

<div align="center">

APPENDIX II

LEAST FAVORABLE CHANNEL

</div>

The result proved below is a generalization of [35, Theorem 2]. In what follows, we shall use $\mathcal{U}_n$, $\mathcal{D}_n$, and $\mathcal{P}_n$ to denote the sets of all $n \times n$ unitary, doubly stochastic, and permutation matrices, respectively.

*Theorem 4:* Consider the nonnegative real numbers $\lambda_k$, $k = 1, \ldots, m$, and $\theta_l$, $l = 1, \ldots, n$, with $m \leq n$, sorted in ascending order. Let the $m \times n$ matrix $\mathbf{\Lambda}$ be such that $\mathbf{\Lambda}(k,k) = \lambda_k^{1/2}$ for $k = 1, \ldots, m$ and $\mathbf{\Lambda}(k,l) = 0$ for $k \neq l$. Denoting the set of all $n \times n$ unitary matrices by $\mathcal{U}_n$ and letting the $n \times n$ matrix $\mathbf{\Theta}$ be given by $\mathbf{\Theta} = \mathrm{diag}\{\theta_l\}_{l=1}^n$, we have

$$\min_{\mathbf{Q} \in \mathcal{U}_n} \mathrm{Tr}\left(\mathbf{\Lambda}\,\mathbf{Q}\,\mathbf{\Theta}\,\mathbf{Q}^H\mathbf{\Lambda}^H\right) = \sum_{k=1}^m \lambda_k\,\theta_{m+1-k}.$$

*Proof:* Straightforward manipulations show that

$$\min_{\mathbf{Q} \in \mathcal{U}_n} \mathrm{Tr}\left(\mathbf{\Lambda}\,\mathbf{Q}\,\mathbf{\Theta}\,\mathbf{Q}^H\mathbf{\Lambda}^H\right) = \min_{\mathbf{Q} \in \mathcal{U}_n} \sum_{k=1}^m \lambda_k \sum_{l=1}^n \theta_l |\mathbf{Q}(k,l)|^2$$

$$\geq \min_{\mathbf{D} \in \mathcal{D}_n} \sum_{k=1}^m \lambda_k \sum_{l=1}^n \theta_l \mathbf{D}(k,l) \tag{119}$$

where $\mathbf{D}$ with $\mathbf{D}(i,j) = |\mathbf{Q}(i,j)|^2$ is doubly stochastic whenever $\mathbf{Q}$ is unitary. The inequality in (119) is a consequence of enlarging the set of admissible matrices, i.e., $\mathcal{U}_n \subset \mathcal{D}_n$. Since the set of doubly stochastic matrices is a compact convex set, a linear function, such as the one in (119), attains its minimum at an extreme point of this set [18, Appendix B]. By Birkhoff's Theorem [18, Theorem 8.7.1], the extreme points of the set of doubly stochastic matrices are the permutation matrices. Hence,

$$\min_{\mathbf{D} \in \mathcal{D}_n} \sum_{k=1}^m \lambda_k \sum_{l=1}^n \theta_l \mathbf{D}(k,l) \geq \min_{\mathbf{P} \in \mathcal{P}_n} \sum_{k=1}^m \lambda_k \sum_{l=1}^n \theta_l \mathbf{P}(k,l)$$

$$= \min_{\mathbf{P} \in \mathcal{P}_n} \sum_{k=1}^m \lambda_k\,\theta_{\pi(k)}$$

$$= \sum_{k=1}^m \lambda_k\,\theta_{m+1-k}. \tag{120}$$

The proof is concluded by noting that permutation matrices also belong to the set of unitary matrices, i.e., $\mathcal{P}_n \subset \mathcal{U}_n$, so that the minimum in (120) is attained with equality. ∎